\documentclass[11pt,letterpaper,]{article}
\usepackage{graphicx}
\usepackage{float}
\usepackage{amsmath}
\usepackage{amsfonts}
\usepackage{amssymb}
\usepackage{subfigure}
\usepackage{epstopdf}
\usepackage{booktabs}

\usepackage[affil-it]{authblk}

\usepackage{mathtools}
\usepackage{marvosym}

\usepackage{xcolor}
\usepackage{footnote}
\usepackage{paralist}
\usepackage[inline]{enumitem}
\usepackage{enumerate}
\usepackage{fancyhdr}
\usepackage{boxedminipage}
\usepackage{algpseudocode}
\usepackage{algorithm}
\usepackage{algorithmicx}
\usepackage{varwidth}
\usepackage{multirow}
\usepackage{lscape}

\usepackage{tablefootnote}
\usepackage{bigfoot}

\usepackage{undertilde}
\usepackage[inline]{enumitem}

\usepackage{setspace}
\usepackage[round]{natbib}

\newcommand{\qed}{\nobreak \ifvmode \relax \else
      \ifdim\lastskip<1.5em \hskip-\lastskip
      \hskip1.5em plus0em minus0.5em \fi \nobreak
      \vrule height0.75em width0.5em depth0.25em\fi}

\newcommand{\bsb}{\boldsymbol}

\newcommand{\bsbX}{{\boldsymbol{X}}}
\newcommand{\bsbx}{{\boldsymbol{x}}}

\newcommand{\bsbY}{{\boldsymbol{Y}}}

\newcommand{\bsbH}{{\boldsymbol{H}}}
\newcommand{\bsbG}{{\boldsymbol{G}}}
\newcommand{\bsbI}{{\boldsymbol{I}}}

\newcommand{\bsbZ}{{\boldsymbol{Z}}}
\newcommand{\bsbP}{{\boldsymbol{P}}}
\newcommand{\bsbQ}{{\boldsymbol{Q}}}

\newcommand{\bsbxi}{{\boldsymbol{\xi}}}
\newcommand{\bsbXi}{{\boldsymbol{\Xi}}}

\newcommand{\bsbD}{{\boldsymbol{D}}}

\newcommand{\bsbV}{{\boldsymbol{V}}}

\newcommand{\bsba}{{\boldsymbol{\alpha}}}
\newcommand{\bsbA}{{\boldsymbol{A}}}

\newcommand{\bsbu}{{\boldsymbol{u}}}

\newcommand{\bsbq}{{\boldsymbol{q}}}

\newcommand{\bsbS}{{\boldsymbol{S}}}
\newcommand{\bsbs}{{\boldsymbol{s}}}

\newcommand{\bsbTh}{{\boldsymbol{\Theta}}}

\newcommand{\bsbmu}{{\boldsymbol{\mu}}}
\newcommand{\bsbnu}{{\boldsymbol{\nu}}}

\newcommand{\bsbv}{{\boldsymbol{v}}}
\newcommand{\bsbB}{{\boldsymbol{B}}}
\newcommand{\bsbz}{{\boldsymbol{z}}}

\newcommand{\bsbDelta}{{\boldsymbol{\Delta}}}

\newcommand{\real}{\mathbb{R}}

\DeclareMathOperator*{\argmin}{argmin}

\DeclareMathOperator*{\card}{card}
\newcommand{\specialcell}[2][c]{%
  \begin{tabular}[#1]{@{\textbf{•}}c@{}}#2\end{tabular}}

\newcommand{\modelname}{\text{SG-PCA}}
\setcitestyle{citesep={,}}

\begin{document}

\title{Sparse Generalized Principal Component Analysis for Large-scale Applications beyond Gaussianity}
\author{Qiaoya Zhang, Yiyuan She}
\affil{Department of Statistics, Florida State University\\Tallahassee, FL 32306-4330}
\date{}
\maketitle

\begin{abstract}
	Principal Component Analysis (PCA) is a dimension reduction technique. It produces {inconsistent} estimators when the dimensionality is moderate to high, which is often the problem in modern large-scale applications where algorithm scalability and model interpretability are difficult to achieve, not to mention the prevalence of missing values.
	While existing sparse PCA methods alleviate inconsistency, they are constrained to the Gaussian assumption of classical PCA and fail to address algorithm scalability issues. %attack
	We generalize sparse PCA to the broad exponential family distributions under high-dimensional setup, with built-in treatment for missing values. Meanwhile, we propose a family of iterative sparse generalized PCA (\text{\text{\modelname }}) algorithms such that despite the non-convexity and non-smoothness of the optimization task, the loss function decreases in every iteration. In terms of ease and intuitive parameter tuning, our sparsity-inducing regularization is far superior to the popular Lasso. Furthermore, to promote overall scalability, accelerated gradient is integrated for fast convergence, while a progressive screening technique gradually squeezes out nuisance dimensions of a large-scale problem for feasible optimization. High-dimensional simulation and real data experiments demonstrate the efficiency and efficacy of \text{\text{\modelname }}.

\end{abstract}
% keyword: unconventional loss function, VS formulation to handle the non-convex constraint ,built-in treatment for missing values,
% sparsity inducing non-convex constraint, introduces quantitle thresholding, group and element-wise sparsity for selection and fast computation, progressive screening - reduced space each time
%	non-convex, nonsmooth overall optimization criterion, used a linearization - optimization technique to gaurantee function value decreasing under such setting,  high dimensionality, quantitle thresholding, easy tuning,

\section{Introduction}

	Suppose an $n\times p$ matrix $\bsbX$ represents a data set with $n$ observations on $p$ variables centered column-wise. In modern statistical applications, both $n$ and $p$ can be quite large especially with $p \gg n$.
	Principal Component Analysis (PCA) is a well-known and popular dimension reduction technique. It \textit{sequentially} searches $r$ ($r \ll p$) leading principal directions along which the projected data points have maximal variance. Equivalently, it can also be realized \textit{jointly} by solving a multivariate low-rank matrix approximation problem
	\begin{gather}	\label{squareErrLoss}
	\min_{\bsbB: rank(\bsbB)\leq r} \| \bsbX - \bsbB\|_F^2,
 	\end{gather}
	with the optimal $\bsbB$ given by $\bsbX \bsbV_r \bsbV_r^T$, where $\bsbV_r = [\bsbv_1, \cdots, \bsbv_r]$ are formed by the top $r$ right singular vectors of $\bsbX$ \citep{eckart1936approximation}. The principal components (PCs) $\{\bsbz_1, \cdots, \bsbz_r\}$ are then given by $\bsbZ = \bsbX \bsbV$ and $\bsbv_i$'s are called the principal loading vectors.
	Clearly, PCs are also the eigenvectors of the sample covariance matrix, which further explains its power in illustrating the variability within the multivariate data in a lower-dimensional space.

%% Challenges

	There are, however, many characteristics to modern large-scale data that regular PCA finds inappropriate to handle. Three specific challenges we want to address in this article are: \begin{inparaenum}[a)] \item the prevalence of non-Gaussian Data, \item the curse of dimensionality as well as \item the existence of missing entries in data arrays\end{inparaenum}.

\subsection{Modern challenges}
% Non-Gaussian Data
	\paragraph{Non-Gaussianity}
	PCA is often employed without addressing any parametric assumption on the original data set $\bsbX$, however, its criterion inherently assumes a Gaussian distribution: the squared-error loss function (\ref{squareErrLoss}) does not make the best sense with, for example, the misclassification error associated with categorical data. The sample covariance matrix, whose eigenvectors are computed as principal components, does not capture all kinds of association either. When the data does not follow a Gaussian distribution, an alternative loss functions might be more proper in measuring the affinity between $\bsbX$ and its low-rank estimate. There are a lot of real-world motivations: Netflix's user rating system is typical of a categorical data type, SNPs data denotes mutation by binary coding, and spam email detection often examines the number of times a flagged word appears.

% Big Data Computation
	\paragraph{Curse of dimensionality}	
	Besides the jeopardy of soundness PCA faces when prompted with an extended range of data types, the method itself also encounters theoretical, practical, and computational challenges when the dimensionality $p$ is high.
		
    Theoretically, the estimated PCs prove to be \textit{inconsistent} if $\lim_{n \rightarrow \infty} p/n$ does not go to zero \citep{johnstone2009consistency, paul2007asymptotics, nadler2008finite}, that is, when the number of variables $p$ is comparable or larger than $n$, which is often the case in applications related to network, genetics and so on. This conclusion manifests a surprising and controversial result: the statistical accuracy of PCA is highly restricted to $p \ll n$ despite that its incentive stems from the challenge of high dimension itself.

    What's worse, when the squared signal-to-noise ratio is less than the converging constant of $p/n$, the estimated principal components can be asymptotically orthogonal to the true ones, essentially containing no authentic information at all! Practically, since each PC $\bsbz_i = \sum_{j = 1}^p \bsbx_j v_{ji}$ still utilizes \textit{all} $p$ dimensions in $\bsbX$ with the loading matrix as weights, there is essentially no hope of interpretability if the entries in $\bsbV$ are mostly nonzero. Therefore, the ideal model in terms of both theoretical consistency and practical selective property, calls for a parsimonious representation of the original dimensions, which corresponds to enforcing sparse nonzero entries in the loading matrix.
	
	The big data era exacerbates the curse of dimensionality from the computational perspective. Even when $p$ is moderately high, many existing algorithms find it difficult to handle such computation complexity. Fast and feasible algorithms need to be developed without sacrificing accuracy.

%	where an unregularized linearization step takes place first to utilize the first-order information conveniently which guarantees function value decreasing, proceeded by a constrained optimization problem under the quadratic loss, to which there exist a handful of desirable properties.
	\paragraph{Missing values}
	One might argue that missing value is not as modern a challenge as its previous peers, but it is for sure a problem that evolves with modern data formats. The Netflix's MoveLens, for example, has a missing percentage of up to 99\% \citep{Koren2008} and a pattern of missingness distinct from traditional problem. If these entries were to be removed or imputed in a conventional fashion, the training of the model is likely to suffer from a great deal of inaccuracy due to missingness. Under such circumstances, it is much desired to develop a novel approach which makes no assumptions on the missing entries and integrate them into the optimization problem itself.\\

%	Sampling theory is mentioned here to reduce the size of workable data matrix while keeping essential information in the reduced dataset\cite{tille2011sampling}.

%	The goal of this paper is therefore to carry out the sparse PCA procedure on data types that assume distributions beyond Gaussian who can also tackle the matrix completion task, thus the three challenges may be addressed simultaneously.

	With these challenges in mind, we examine the existing literature for ideas explored.
	On the topic sparse PCA, there are quite a few iterative estimating algorithms such as \citet{jolliffe2003modified}, \citet{Shen&Huang}, \citet{Witten}, \citet{Zou&Hastie&Tibshirani}, among others, where each column of the loading matrix is sequentially retrieved. Despite of its simple formulation and the resulting nested spaces, sequential techniques lack joint optimality and joint orthogonality when multiple features are desired. More importantly, these methodologies are proposed under Gaussian assumption.
	Generalizations of PCA to fit various non-Gaussian data types are seen mostly in the machine learning community targeted on specific applications, such as \citet{kramer1991nonlinear}, \citet{hofmann1999probabilistic}, \citet{blei2003latent}, \citet{schein2003generalized}, \citet{de2006Principal}, \citet{linting2007nonlinear}. \citet{collins2001generalization} generalizes PCA to the exponential family. However, it discusses the rank-one case only under a large $n$ background. In the statistical literature, perhaps the most similar work to ours is logisticPCA by \citet{lee2010sparse} and \citet{lee2013coordinate}---both utilizing $\ell_1$ regularization for sparsity, whose parameter is tuned by a greedy sequential grid search which may result in a suboptimal estimator. The criterion is established for binary data in specific, a subset of our target. Missing values are discussed, but proposed to handle in a conventional imputation manner. More details of logsiticPCA will be given in Section \ref{sec:simulation}. 	
	None of the aforementioned works addresses algorithm scalability when $p$ is high and computationally demanding. While sparsity lies in the heart of these methods, it is often enforced rank by rank such that no overall selection power is guaranteed unless the rank level is very low. In sum, there is a scarcity of existing literature that compares to the scope and scale of our work.

\subsection{An outline}
	We propose a novel low-rank multivariate data approximation method called \text{\modelname }.
	\text{\modelname } establishes a universal framework for any exponential family distribution under the high-dimensional setting. We launch from a low-rank data approximation perspective and propose a joint rank-$r$ algorithm with built-in treatment for missing values. To achieve this goal, we solve a non-smooth optimization criterion with non-convex rank, sparsity and orthogonality regularization.
	Matrix decomposition eases the low-rank condition in the optimization criterion. {Element-wise} and {group-wise} sparsity constraints are studied and differentiated. A Stiefel manifold optimization problem is simplified by an iterative process with some MM (majorization minimization) flavor.
	As a result, we are able to recover the loading space with rank reduction and dimension selection all in one step, while rectifying the inconsistency issue brought by high dimensionality.
	Our formulation treats missing values inherently as a part of the criterion.
	For fast computation, accelerated gradient methods are incorporated in \text{\modelname } to promote algorithm efficiency. Last but not least, we develop what is  essential to the feasibility of large-scale applications---a progressive screening scheme which throws away nuisance dimensions gradually, providing a smaller problem size each round.

	%	Note also that sparse on an element-wise scope alone is not sufficient to eliminate nuisance dimensions - sparse columns of $\bsbB$ is necessary for selectable dimension.

%	Let $\bsbB$ denote the low-rank representation of the original data $\bsbX$. To cope with large $p$, the double sparsity and low-rank regularization on $\bsbB$ is extremely complex. The non-convex rank constraint can be eliminated simply by rewriting $\bsbB = \bsbV \bsbS$, where $\bsbV$ is the left singular vectors of $\bsbX$, and $\bsbS$ the scaled loading matrix. Hence we constrain $\bsbS$ sparse and $\bsbv_i$'s orthonormal instead such that the desired sparse PCA may be achieved.

%	
%	We first choose to realize it through retraction method via manifold optimization. To enhance convergence speed, we discuss the step size of the manifold optimization algorithm: a non-monotone search scheme \citep{barzilai1988two} is applied to achieve superlinear convergence rate. Two other schemes designed inspired by the idea of a quadratic-surrogate function in the place of the original non-quadratic loss, such that procrustes rotation can be used to solve the problem on $\bsbV$.

	The rest of the paper is organized as below. In Section \ref{sec:body}, we formulate the criterion for non-Gaussian data with missing values and describe the necessity and approach to enforce sparse loading vectors. Section \ref{sec:opt} gives the \text{\text{\modelname }} algorithms to solve the rank-$r$ problem under both sparsity and orthogonality constraints and incorporate accelerated gradient methods, a line search scheme, and a powerful progressive screening strategy to enhance algorithm efficiency and scalability. Simulation studies and real data applications are given in Section \ref{sec:experiment}.

\section{Sparse generalized PCA}	\label{sec:body}	
	This section is dedicated to formulate and layout the \text{\modelname } procedure in every detail. We start with deriving a suitable criterion, proceed to discussion of the penalties and constraints used for regularization and the missing data issues.
\subsection{The regularized log-likelihood criterion}
	
	Suppose our data is stored in an $n \times p$ matrix $\bsbX$ that follows some underlying distribution. Low-rank data approximation methods search for the projection $\bsbB$ of the data to an $r$-dimensional ($r < p$) subspace such that the loss of information $L(\bsbX, \bsbB)$ is minimized.
	Under the Gaussian assumption, sparse PCA finds $\bsbB$ by maximizing a sum-of-squares criterion $\|\tilde{\bsbX} - \tilde{\bsbu}\tilde{\bsbv}^T \|_F^2 + P_\lambda(\bsbv)$ sequentially, the data matrix deflated each time getting a pair of $(\tilde{\bsbu}, \tilde{\bsbv})$ \citep{Shen&Huang}. The sparsity regularization is enforced on $\tilde{\bsbv}$ to obtain an estimated loading vector. Therefore, sparse PCA has a closed-form update under the simple quadratic loss.
	
	However, our data of interest encompass all distributions in the exponential family, for which the sum-of-squares criterion may fail. The negative log-likelihood is an intuitive alternative for such non-Gaussian data types.
	
	This extension of loss function resembles that of the generalized linear models (GLMs). GLMs deal with non-Gaussian response variables for the lack of linear relationship between the predictors and the response.	Let $\bsbmu = E(\bsbX)$ and $g(\cdot)$ denote the canonical link function, then $\bsbTh = g(\bsbmu)$ is the natural parameter. We assume that all the $x_{ij}$'s are independent given the low-rank structure in $\bsbTh$ and write the negative log-likelihood function in matrix form:
	\begin{align}\label{NegLogLike}
	-\mathit{l}(\bsbTh|\bsbX) = -\langle \bsbX , \bsbTh \rangle + \langle \bsb1_n \bsb1_p^T, b(\bsbTh)\rangle -\log(h(\bsbX)),
	\end{align}
	where the matrix component-wise inner product is defined as $\langle \bsbA, \bsbB \rangle =  tr(\bsbA^T \bsbB)$, the term $\log(h(\bsbX))$ is treated as a constant, thus omitted during optimization. Note that the derivative of the log partition function $b'(\cdot)$ is equal to the inverse of the canonical link function $g^{-1}(\cdot)$. The negative log-likelihood of a Gaussian $\bsbX$ returns equation (\ref{squareErrLoss}), the squared-error loss. Table \ref{tb:functions} gives a list of functions of interest with respect to some commonly used exponential family distributions, where $\mu$ and $\theta$ stand for the mean and natural parameter in general.

	\begin{table}[H]
		\centering	
%		\begin{minipage}{\linewidth}
%		{\renewcommand{\arraystretch}{0.92}%
%		\resizebox{\linewidth}{!}{
		\begin{tabular}{ccccc}
			\toprule
			Distribution & $g(\mu)$ & $b'(\theta) = g^{-1}(\theta)$ & $b(\theta)$ & $ b''(\theta)$ \\
			\midrule\midrule
			Gaussian 	 & $\mu$    & $\theta$   					& $\frac{\theta^2}{2}$ & 1\\ \midrule
			\specialcell{Bernoulli\\ Binomial\\  Multinomial}
						& $\log \frac{\mu}{1-\mu}$ & $\frac{e^{\theta}}{1+e^{\theta}}$ & $\log (1+e^{\theta})$& $\frac{1}{e^\theta+e^{-\theta}+1}$\\ \midrule
			Poisson     & $\log{\mu}$ & $e^\theta$ & $ e^\theta$ &$e^\theta$ \\ \midrule
			\specialcell{Exponential\\Gamma}
						& $-\frac{1}{\mu}$ &$ -\frac{1}{\theta}$ & $-\log (|\theta|)$ & $\frac{1}{\theta^2}$ \\
			\bottomrule
		\end{tabular}
%	}
%	}
%		\end{minipage}
		\caption{A list of functions of interest with respect to various distributions}\label{tb:functions}
	\end{table}
	
	To represent the natural parameter in a low-rank fashion, we rewrite $\bsbTh = \bsb1_n \bsba^T +\bsbV \bsbS^T $, where $\bsb1_n$ is a column vector of $1$'s and $\bsba$ the intercept vector. Here we require $\bsbV \in \mathbb{O}^{n \times r}= \{\bsbA \in \real^{p \times r}|\bsbA^T \bsbA = \bsbI_{r \times r}\} $. $\bsbS \in \real^{p \times r}$ gives the principal loading matrix but does not necessarily have orthogonal columns. Instead, we require $\bsbS$ to be sparse. Such a $\bsbV\bsbS^T$ setup prepares the objective function for regularization {ease}.
	
	The objective function minimizes the regularized negative log-likelihood:
	\begin{gather}\label{SGPCA}
	\begin{split}
	-\langle \bsbX, \bsb1_n \bsba^T +\bsbV \bsbS^T\rangle + \langle\bsb1_n \bsb1_p^T, b(\bsb1_n \bsba^T + \bsbV \bsbS^T)\rangle  + P ( \bsbS; \lambda)\\
	\text{subject to } \bsbV^T \bsbV = \bsbI_{r \times r}	
	\end{split}
	\end{gather}
	where $ P(\bsbS; \lambda)$ denotes a sparsity-inducing regularization with $\lambda $ as its parameter.
	
	This criterion is applicable to a variety of large-scale applications for its wide assumptions on distributions.
	The entire low-rank approximation can also be derived as a whole, preserving joint optimality as compared to the sequential fashion.

\subsection{$\bsbS$ for sparsity}
	Since $\bsbS$ denotes the principal loading matrix, the PCs are written as $\bsbz_i = \bsbX \bsbs_i$. When the dimension $p$ is high, relative to the sample size $n$, it is necessary for $\bsbS$ to have {sparse} nonzero entries to ensure consistency \citep{johnstone2009consistency}.	
	Furthermore, the scope of dimension reduction can and should be improved by the addition of sparsity constraints.
	Dimension reduction traditionally refers to \textit{rank} reduction in the context of PCA. Geometrically it is a projection of the observed data points to a lower $r$-dimensional subspace.
	What promotes model parsimony is the elimination of nuisance dimensions along with rank reduction, leading to sparse representation in the PCs.
	
	On the other hand, enforcing element-wise sparsity does not yield the smallest subset of variables for principal components unless $r$ is extremely low. To remove an entire column of $\bsbX$ in the construction of all PCs given by $\bsbZ = \bsbX \bsbS$, introducing \textit{row-sparsity} in $\bsbS$, i.e., $P(\bsbS; \lambda) = \sum_{i = 1}^p P(\|\bsbs_i\|_2; \lambda)$ (with some abuse of notation), would do the trick. It is easy to see that the row-sparsity in $\bsbS$ corresponds to the column sparsity in $\bsbB = \bsbV\bsbS^T$, verified due to the orthoganality of $\bsbV$. It is essential for efficiency and selectivity of the algorithm and for the overall interpretability of the results.
	We will blend the two types of sparsity on the loading matrix for fast computation, with row sparsity employed in a screening step to reduce $p$ to some dimension $d$ ($d < p$) prior to the element-wise sparsity pursuit in each individual loading vector.
	
	%%%%%%%%%% sparsity penalty%%%%%%%%%%%%%

	There are abundant choices of sparsity-inducing penalty functions $P(\cdot; \cdot)$ in $ \sum_{i, j} P(|s_{ij}|; \lambda )$ and $\sum_{i = 1}^p P(\|\bsbs_i\|_2; \lambda) $.
	The $\ell_1$ penalty \citep{tibshirani1996lasso} is most popular among the sparse PCA literature, but it suffers from inconsistency and biased estimation \citep{zhao2006model,zhang2010nearly} especially when predictors are correlated.
	To alleviate those issues, other non-convex approximations of the ideal non-convex $\ell_0$ ($\|\bsbS\|_0 = \sum_{i,j} 1_{s_{ij} \neq 0} $) penalty such as $\ell_p$ $( 0 < p <1 )$ penalty, SCAD \citep{fan2001variable} and capped $\ell_1$ \citep{zhang2010analysis} are proposed. However, we propose to use the $\ell_0$ itself for its appealing properties in sparsity promotion, because it can directly limit the cardinality of nonzero elements/rows in the loading matrix hence encourage accurate selection.
	Moreover, the \textbf{constraint} form of $\ell_0$ is yet preferred over the penalty form due to its tuning ease. While the grid search process of $\lambda$ is traditionally cumbersome, tuning for the constraint form $\sum_{i,j} 1_{s_{ij} \neq 0}/(pr) \leq q_e$ is intuitive and easy: $q_e$ serves as the upper bound for percent of nonzero entries in the loading matrix, thus the selection accuracy should remain sound as long as parameter $q_e$ is larger than the true value. In fact, it corresponds to a rank-constrained screening problem described in \cite{selrrr}.

	%%%%%%%%%%%
	Note that orthogonality is only imposed on $\bsbV$. While this makes $\bsbV \bsbS^T$ a general representation for any matrix $\bsbB$ of rank no higher than $r$---a convenient translation of the non-convex low-rank constraint, it implies that the obtained sparse PCs are not decorrelated as in ordinary PCA.
	However, the loss of orthogonality is generally considered a price that sparse PCA has to pay \citep{Zou&Hastie&Tibshirani,Shen&Huang,Witten}.

\subsection{Handling missing values }

	We propose a masking approach for efficient handling of missing entries.
	Conventionally, it is a common practice to impute or simply remove missing entries before training the model, both holding assumptions on the missing entries thereby introducing additional inaccuracy into the training process. On the contrary, we do not assume any prior knowledge on the missing entries, rather, we mask them as unknown information such that their contribution to the loss function is not taken into consideration. This is made possible by approximating the low-rank and sparse structure of the data, such that masking some amount of missing values does not interfere with matrix recovery.
	Perhaps it is interesting to note that our masking technique for missing data is deeply connected and naturally applicable to matrix completion type of problems (see examples in \cite{candes2009exact}), where only an extraordinary small fraction of data is observed and {imputation} of the missing values is precisely the goal.
	Instead of writing the problem as a summation in a subset, we will develop a  \textit{multivariate} approach that utilizes matrix-wise operations, making it easier in implementation.   The masking method introduces minimal cost to the computational algorithm even as the dimension of the problem is rocket high, and it is entirely integrated into the estimation process.

	Let $\Omega$ denote the index set of all {available} observations, intuitively the optimization criterion \eqref{SGPCA} is only evaluated when $\bsbX \in \Omega$. However, for computation efficiency and ease in analysis, we prefer the following formulation. 	
	Define the masking matrix $\bsbH=[h_{ij}]$ such that
	\begin{equation}
	h_{ij}=\begin{cases}
	1 & \text{if } (i,j)\in \Omega\\
	0 & \text{if } (i,j) \in \Omega^C
			\end{cases}
   	\end{equation}	
	Throughout the paper we use the Hadamard Product $\circ$ to denote the element-wise matrix multiplication: $\forall \bsbX$ where $dim(\bsbX) = dim(\bsbH)$,
	$$\bsbH \circ \bsbX = [h_{ij} \cdot x_{ij}].$$
	From here on, our attention shifts to the loss function $f^{m}$ below:
	\begin{gather}
		-\langle {\bsbH \circ \bsbX}, \bsb1_n \bsba^T+\bsbV \bsbS^T \rangle
		+ \langle \bsbH, b(\bsb1_n \bsba^T+\bsbV \bsbS^T) \rangle
	\end{gather}	
	
%	With the change of objective function comes the updates of gradients.
%	A deduction of gradient with respect to $\bsbV$ is given below as an example.
%	Let $\bsbDelta^{\cdot 2}$ denote $[\delta_{ij}^2]$, the point-wise squared matrix.
%	\begin{align*}
%	f^{m}(\bsbV)=-\langle \tilde{\bsbH \circ \bsbX}, \bsb1_n \bsba^T+\bsbV \bsbS^T \rangle
%		+ \langle \bsbH, b(\bsb1_n \bsba^T+\bsbV \bsbS^T) \rangle\\
%	\end{align*}
%	Then
%	\begin{align*}
%	f^{m}(\bsbV + \bsbDelta) &=  -\langle \tilde{\bsbX}, \bsb1_n \bsba^T+(\bsbV +\bsbDelta) \bsbS^T \rangle
%							+ \langle \bsbH, b(\bsb1_n \bsba^T+(\bsbV +\bsbDelta) \bsbS^T) \rangle \\
%						 & = -\langle \tilde{\bsbX}, \bsb1_n \bsba^T+\bsbV \bsbS^T \rangle
%						 	-\langle \tilde{\bsbX}, \bsbDelta \bsbS^T \rangle
%						 	+ \langle \bsbH, b(\bsb1_n \bsba^T+\bsbV \bsbS^T)\\
%						 	&+b'(\bsb1_n \bsba^T+\bsbV \bsbS^T)\circ (\bsbDelta \bsbS^T)
%						 	+ (\bsbDelta \bsbS^T) \circ (\bsbDelta \bsbS^T) \rangle \\
%						 & = f^{m}(\bsbV)
%						 -\langle \tilde{\bsbX}\bsbS, \bsbDelta
%						  \rangle
%						 + \langle \bsbH \circ b'(\bsb1_n \bsba^T+\bsbV \bsbS^T),\bsbDelta \bsbS^T \rangle
%						 + \bsbDelta^{ \cdot 2}{\bsbS^{\cdot 2}}^T
%	\end{align*}
%	Therefore
%	\begin{align*}
%	\nabla_\bsbV f^{m} =  -\tilde{\bsbX}\bsbS
%						 				+ ( \bsbH \circ b'(\bsb1_n \bsba^T+\bsbV \bsbS^T)) \bsbS
%	\end{align*}
%	Similarly, the masked gradients can be calculated as below:

	The gradients of the masked loss function can be calculated (details omitted):
	\begin{align}\label{func:gradients}
	\begin{cases}
		&{\bsbG}_{\bsbS}= - {(\bsbH \circ \bsbX)}^T \bsbV + (\bsbH \circ g^{-1}(\bsb1_n \bsba^T + \bsbV \bsbS^T))^T \bsbV\\
		&{\bsbG}_{\bsbV}= - {(\bsbH \circ \bsbX)} \bsbS + (\bsbH\circ g^{-1}(\bsb1_n \bsba^T + \bsbV \bsbS^T)) \bsbS\\
		&{\bsbG}_{\bsba} = -{(\bsbH \circ \bsbX)}^T \bsb1_n + (\bsbH\circ g^{-1}(\bsb1_n \bsba^T + \bsbV \bsbS^T))^T \bsb1_n
	\end{cases}	
	\end{align}

	For simplicity, we use ${\bsbX} = \bsbH \circ \bsbX$ hereafter to demonstrate the masked observed data matrix.

%		& \tilde{\bsbG}_{\bsbTh}= -\tilde{\bsbX}^T + \bsbH \circ g^{-1}(\bsbTh)
%	Note that incorporating the missing values alters the optimization for $\bsba$. In cases where all entries are observed, $\bsba^{[k]}$ has closed-form solutions; but with the mask no closed-form solution may be achieved for $\bsba^{[k]}$, an iterative procedure is employed instead.

\section{An iterative algorithm}\label{sec:opt}

\subsection{A surrogate function}
	A natural idea to solve optimization problem \ref{SGPCA} is to utilize a block coordinate descent (BCD) \citep{tseng2001convergence} algorithm where $\bsba$, $\bsbS$ and $\bsbV$ are updated alternatively. While $\bsba^{[k]}$ has a closed-form solution when all entries are observed and the distribution is Gaussian, in the existence of missing values or under GLM setting, $\bsba^{[k]}$ has no explicit form.
	
	The bigger challenge lies in optimizing $\bsbV$ while holding $\bsba$ and $\bsbS$ fixed. Although the objective function is smooth in $\bsbV$, the {unitary constraint} $\bsbV^T \bsbV = \bsbI_{p \times p} $ is non-convex and non-smooth. One may treat the update of $\bsbV$ as a constrained optimization problem with quite a few Lagrangian multipliers, but it is awkward and slow in computation. 	
	The problem is better phrased as a Stiefel manifold optimization one, for which packages are already available \citep{wen2013feasible,she2015robust}. That is what we first try. We have implemented and tested the manifold optimization algorithm for \text{\modelname } only to find it to be a valid yet rather expensive method when the problem size is big. With no need to call any external package we develop a new algorithm with some MM (majorization minimization) flavor in its employment of a surrogate function.

	Consider the optimization problem in the element-wise sparse case as an instance. We minimize the objective function
	\begin{align}
	\begin{split}
	\min_{\bsba, \bsbV, \bsbS}f = -\langle {\bsbX}, \bsb1_n \bsba^T+\bsbV \bsbS^T \rangle
		+ \langle &\bsbH, b(\bsb1_n \bsba^T+\bsbV \bsbS^T) \rangle+ P(\bsbS; \lambda)\\
			&\text{subject to } \bsbV^T \bsbV = \bsbI_{p \times p},
	\end{split}
	\end{align}
	where $ l(\bsba, \bsbV, \bsbS) = -\langle {\bsbX}, \bsb1_n \bsba^T+\bsbV \bsbS^T \rangle
	+ \langle \bsbH, b(\bsb1_n \bsba^T+\bsbV \bsbS^T) \rangle$ denotes the loss.
	
	Solving this non-quadratic loss function with a non-convex orthogonality constraint is rather difficult, for the prerequisites of the convenient candidate approaches such as Procrustes rotation are deprived. We seek to utilize a quadratic loss through deriving a surrogate function.
	
%	first proceeding one step along the gradient of $l(\bsbTh) = -\langle \bsbX, \bsbTh \rangle+\langle \bsbH, b(\bsbTh)\rangle$: $\bsbTh^{[k]}=\bsbTh^{[k-1]}-\tau_k \nabla_\Theta l(\Theta^{[k-1]})$, whose theoretical step size $\tau_k$ will be derived under applicable circumstances later. Then we optimize the surrogate function through an inner loop.

%	Let $\bsbb$ denote the parameter of interest, function $h$ is called the majorizing function if
%	\begin{align}
%	h(\bsbb, \bsbb') \geq f(\bsbb) \text{   } \forall \bsbb  \text{   and   } h(\bsbb, \bsbb) = f(\bsbb)
%	\end{align}
	
	Concretely, given the $(k-1)^{th}$ step estimates $\bsba^{[k-1]}, \bsbS^{[k-1]}, \bsbV^{[k-1]}$, define $\bsbTh^{[k-1]}$ as a function of $\bsba^{[k-1]}$, $\bsbV^{[k-1]}$, $\bsbS^{[k-1]}$:
	$$\bsbTh^{[k-1]} = \bsb1_n {\bsba^{[k-1]}}^T +\bsbV^{[k-1]} {\bsbS^{[k-1]}}^T. $$
	Linearization will be applied at $\bsbTh^{[k-1]}$ instead---a similar idea we used in \cite{she2010reduced} to deal with singular-value penalized vector GLMs. Define
	\begin{align*}
	h(\bsbTh^{[k]}, \bsbTh^{[k-1]}) =& l(\bsbTh^{[k-1]})
			+ \langle \nabla_\Theta l(\bsbTh^{[k-1]}), \bsbTh - \bsbTh^{[k-1]}\rangle
			+ \frac{\rho_k}{2}\|\bsbTh - \bsbTh^{[k-1]}\|_F^2 \\
			&+P(\bsbS;\lambda)\\
			= & l(\bsbTh^{[k-1]})
			+ \langle  -\bsbX + \bsbH \circ g^{-1}(\bsbTh^{[k-1]}) , \bsbTh - \bsbTh^{[k-1]}\rangle
			+ \\&\frac{\rho_k}{2}\|\bsbTh - \bsbTh^{[k-1]}\|_F^2 +P(\bsbS;\lambda)
	\end{align*}
	as our surrogate function. Then the $k^{th}$ iterate is given by
	\begin{align}
	(\bsba^{[k]}, \bsbS^{[k]}, \bsbV^{[k]})
	=& \argmin_{\bsba, \bsbS, \bsbV\in \mathbb{O}^{n \times r}} h(\bsba, \bsbS, \bsbV; \bsba^{[k-1]}, \bsbS^{[k-1]}, \bsbV^{[k-1]}). \label{func:kIter}
	\end{align}
	
	It is clear that $h$ satisfies $h(\bsba^{[k-1]}, \bsbS^{[k-1]}, \bsbV^{[k-1]}; \bsba^{[k-1]}, \bsbS^{[k-1]}, \bsbV^{[k-1]}) = f(\bsba^{[k-1]}, \bsbS^{[k-1]}, \bsbV^{[k-1]})$. Suppose $\rho_k$ is chosen such that
	\begin{align}
	h(\bsba^{[k]}, \bsbS^{[k]}, \bsbV^{[k]}; \bsba^{[k-1]}, \bsbS^{[k-1]}, \bsbV^{[k-1]}) \geq f(\bsba^{[k]}, \bsbS^{[k]}, \bsbV^{[k]})\label{func:suroIneq}
	\end{align}
	This can be realized by setting a large enough $\rho_k$ based on Taylor expansion, as detailed in Section \ref{sec:stepsize}. The objective function value is guaranteed nonincreasing throughout the iteration, as demonstrated by the sequence below:
	\begin{align*}
	f(\bsba^{[k]}, \bsbS^{[k]}, \bsbV^{[k]})
	&\leq h(\bsba^{[k]}, \bsbS^{[k]}, \bsbV^{[k]}; \bsba^{[k-1]}, \bsbS^{[k-1]}, \bsbV^{[k-1]}) \\
	&\leq h(\bsba^{[k-1]}, \bsbS^{[k-1]}, \bsbV^{[k-1]}; \bsba^{[k-1]}, \bsbS^{[k-1]}, \bsbV^{[k-1]}) \\
	&= f(\bsba^{[k-1]}, \bsbS^{[k-1]}, \bsbV^{[k-1]}).
	\end{align*}
	
	Clearly the second inequality does not depend on the optimality of $(\bsba^{[k]}, \bsbS^{[k]}, \bsbV^{[k]})$ in equation \eqref{func:kIter}. One can use iterative methods which approximates the solution to solve \eqref{func:kIter}, as detailed below. Problem \eqref{func:kIter} can be rewritten as:
	\begin{align*}
	(\bsba^{[k]}, \bsbS^{[k]}, \bsbV^{[k]})
	 =& \argmin_{\bsba, \bsbS, \bsbV\in \mathbb{O}^{n \times r}}
	\langle \nabla_\Theta l(\bsbTh^{[k-1]}), \bsb1_n \bsba^T +\bsbV\bsbS^T - \bsbTh^{[k-1]}\rangle\\
	&+ \frac{\rho_k}{2}\|\bsb1_n \bsba^T +\bsbV\bsbS^T - \bsbTh^{[k-1]}\|_F^2 +P(\bsbS;\lambda)\\
	 =& \argmin_{\bsba, \bsbS, \bsbV\in \mathbb{O}^{n \times r}}
	\frac{1	}{2} \|\bsb1_n \bsba^T +\bsbV\bsbS^T - \bsbXi^{[k]}\|_F^2 + \frac{1}{\rho_k}P(\bsbS; \lambda),
	\end{align*}
	where
	\begin{align}
	\bsbXi^{[k]} = \bsbTh^{[k-1]} + \frac{1}{\rho_k}(\bsbX - \bsbH \circ g^{-1}(\bsbTh^{[k-1]})).\label{func:DefineXi}
	\end{align}
	For notational simplicity, we write the problem as
	\begin{align}
			\begin{split}
			\min_{\bsba, \bsbV, \bsbS} \tilde{f} = \frac{1}{2} \|\bsb1_n \bsba^T + \bsbV \bsbS^T - \bsbXi^{[k]}\|_F^2 + {P(\bsbS; \lambda_{\rho_k})}\\
			\text{subject to } \bsbV^T \bsbV = \bsbI_{r \times r},\label{func:surogate}
			\end{split}
	\end{align}
	where $\lambda_{\rho_k}$ is defined such that $P(t; \lambda)/\rho_k = P(t; \lambda_{\rho_k})$, $\forall t \in \mathbb{R}$.
	The quadratic problem is much simpler than the initial criterion. As illustrated below, BCD can be easily applied, leading to an inner loop from line \ref{Alg1:InnerStart} to \ref{Alg1:InnerEnd} in Algorithm \ref{Alg:ThetaAlt}.

	\paragraph{$\bsba$-optimization}		
	For the $t^{th}$ inner iteration step, $\bsba$ has a closed-form solution: let $\nabla_\bsba \tilde{f}  = 0$, then $(\bsbXi^{[k]}-\bsb1_n \bsba^T - \bsbV \bsbS^T)^T \bsb1_n  = 0$, that is
	$ \bsba \bsb1_n^T \bsb1_n = {\bsbXi^{[k]}}^T \bsb1_n -\bsbS \bsbV^T \bsb1_n$, therefore
	${\bsba}^{opt} = \frac{1}{n}({\bsbXi^{[k]}}^T - \bsbS^{[t-1]} {\bsbV^{[t-1]}}^T)\bsb1_n
	$.
			
	\paragraph{$\bsbS$-optimization}
	To solve the $\bsbS$-optimization problem for a general penalty function $P(t; \lambda)$, we propose to use the thresholding rule based $\Theta$-estimators, because multiple penalty functions often correspond to one threshold rule.
	Thresholding-based iterative selection procedure (TISP) \citep{she2009thresholding} can be used to solve a $P$-penalized problem for any $P$ associated with a thresholding rule (an odd, unbounded monotone shrinkage function) \citep{she2011outlier}. According to \cite{she2012iterative}, $\Theta$-estimator is linked to a general penalty function $P(t;\lambda)$ by
	\begin{gather}
	P(t;\lambda) - P(0;\lambda) = \int_{0}^{|t|} (\sup\{ s: \Theta(s;\lambda) \leq u \} - u)d u + q(t; \lambda)
	\end{gather}
	for some nonnegative $q(\cdot; \lambda)$ such that $q(\Theta(s; \lambda)) = 0 $ for all $s$. This conclusion is valid for any thresholding rule, so through $\Theta$-estimators, we can handle all popular penalties including but not limited to the aforementioned $\ell_1$, SCAD, and $\ell_p$ ($0<p<1$), making this technique universally applicable.
		
	Recall that we advocate the use of $\ell_0$ \textit{constraint}
	$$ \|\bsbS\|_0/(pr) \leq q_e$$ in place of the $\ell_0$ penalty $ \lambda\|\bsbS\|_0 $ for tuning ease. The modified algorithm guarantees the nonincreasing of function value. TISP can be nicely adapted to solve the problem by employing a quantile {thresholding} rule $\Theta^\#$. The quantile thresholding rule is a special case of hard thresholding which correspond to the $\ell_0$ penalty.
	Let $\Theta^\#(\bsbS; q_e)$ be the element-wise quantile threshold function, then
			\begin{gather}\label{thres:element}
			\Theta^\#(\bsbS; q_e) =
			\begin{cases}
			0 & |s_{ij}| \leq \lambda_e \\
			s_{ij} & |s_{ij}| > \lambda_e,
			\end{cases}
			\end{gather}
	where $\lambda_e$ is the $(1-q_e)^{th}$ quantile of the $|s_{ij}|'s$.
	The tuning parameter $q_e$ has a definite range ($0<q_e<1$), which serves as an upper bound for the nonzero percentage hence is much easier to interpret.
	%%%%%%%%%%%%%%%%%%%%%%%%%%%%%%%%	
	Thus, the closed-form solution for $\bsbS$ associated with \eqref{func:surogate} given $\bsba$ and $\bsbV$ is
%			$
%			\bsbS^{[t]} = ({\bsbTh^{[k]}}^T - \bsba^{[t]} \bsb1_n^T)\bsbV^{[t-1]}
%			$.
			$$\bsbS^{opt} = \Theta^\#(({\bsbXi^{[k]}}^T-\bsba\bsb1_n^T)\bsbV;q_e).$$
	See \cite{she2012iterative} for further details.
	\paragraph{$\bsbV$-optimization}		
	The optimization problem given $\bsba$ and $\bsbS$ with respect to $\bsbV$ becomes
	\begin{gather}
	\begin{split}
	\min_\bsbV \|{\bsbXi^{[k]}}^T-\bsba \bsb1_n^T -\bsbS\bsbV^T\|_F^2\\
	\text{ subject to }	 \bsbV^T \bsbV = \bsbI_{r \times r}.
	\end{split}
	\end{gather}
	This can be identified as a Procrustes rotation problem, realizable through computing the Singular Value Decomposition of $({\bsbXi^{[k]}} - \bsb1_n {\bsba}^T)\bsbS = \bsbP \bsbD \bsbQ^T$. The optimal $\bsbV^{opt} = \bsbP \bsbQ^T$.

	The full optimization process is given in details as Algorithm \ref{Alg:ThetaAlt}.

	\begin{algorithm}[t!]
	\caption{The \text{\modelname } Algorithm}\label{Alg:ThetaAlt}
	\textbf{Input}:
	$\bsbX \in \real^{n \times p}$;
	$r$: the desired rank;
	$M_{out}$/$M_{in}$: the maximum outer/inner iteration number; $\varepsilon_{out}/\varepsilon_{in}$: inner and outer error tolerance;
	and the initial estimates $\bsba^{[0]} \in \real^{p \times 1}$,
	$\bsbV^{[0]} \in \mathbb{O}^{n \times r}$,
	$\bsbS^{[0]} \in \real^{p \times r}$.
	\begin{algorithmic}[1]
	\State $k \gets 0$;
	\State $\bsbTh^{[0]} = \bsb1_n \bsba^{[0]}+\bsbV^{[0]}{\bsbS^{[0]}}^T $;
	\Repeat
%			\State $\bsbTh^{[k]}=\bsb1_n {\bsba^{[k]}}^T + \bsbV^{[k]} {\bsbS^{[k]}}^T$, $k = k+1$
			\State $k \gets k+1$;
			\State $\bsbXi^{[k]} = \bsbTh^{[k-1]} + \frac{1}{\rho_k}{(\bsbX - \bsbH \circ g^{-1}(\bsbTh^{[k-1]}))}$;
			\State $t \gets 0$;
			\State Initialize $\tilde{\bsba}^{[0]} \gets \bsba^{[k-1]}$,
					$\tilde{\bsbS}^{[0]} \gets \bsbS^{[k-1]}$,
					$\tilde{\bsbV}^{[0]} \gets \bsbV^{[k-1]}$;
			\Repeat \label{Alg1:InnerStart}
				\State $t \gets t+1$;
				\State $\tilde{\bsba}^{[t]}\gets\frac{1}{n} (\bsbXi^{[k]} - {\tilde{\bsbV}^{[t-1]}} {\tilde{\bsbS}^{[t-1]}}{}^T)^T \bsb1_n$;
				\State ${\tilde{\bsbS}^{[t]}} \gets \Theta^\#(({\bsbXi^{[k]}}^T-\tilde{\bsba}^{[t]} \bsb1_n^T){\tilde{\bsbV}^{[t-1]}};q_e)$;\label{Alg1:SUpdate}
				\State
			\begin{minipage}{\linewidth}
				Compute SVD of $(\bsbXi^{[k]}-\bsb1_n {\tilde{\bsba}^{[t]}}{}^T) {\tilde{\bsbS}^{[t]}} = \bsbP \bsbD \bsbQ^T$, set ${\tilde{\bsbV}^{[t]}} \gets \bsbP \bsbQ^T$;	
			\end{minipage}
			\Until {$t \geq M_{in}$ or changes in $\tilde{\bsba}$, $\tilde{\bsbS}$, $\tilde{\bsbV}$ no bigger than $\varepsilon_{in} $ }\label{Alg1:InnerEnd}
			\State \begin{varwidth}[t]{\linewidth}
					 ${\bsba}^{[k]} \gets \tilde{\bsba}^{[t]}$,
						  ${\bsbS}^{[k]} \gets \tilde{\bsbS}^{[t]}$,
						  ${\bsbV}^{[k]} \gets \tilde{\bsbV}^{[t]}$;
				   \end{varwidth}
			\State 	${\bsbTh}^{[k]}\gets\bsb1_n{{\bsba}^{[k]}}^T +{{\bsbV}^{[k]}} {{\bsbS}^{[k]}}^T$;
	\Until{$k \geq M_{out}$ or ($\|\bsbTh^{[k]}-\bsbTh^{[k-1]}\|_{\max} \leq \varepsilon_{out}$ $\&$ $|f^{[k]}-f^{[k-1]}| \leq \varepsilon_{out}$)}\\
	\Return $\bsba^{[k]}$,
	$\bsbV^{[k]}$,
	$\bsbS^{[k]}$.
	\end{algorithmic}
\end{algorithm}

	Although for simplicity of presentation, the main algorithm is illustrated with the element-wise form, group sparsity regularization can be used for screening and selection purposes and is also interesting to us.
	In particular, for the group $\ell_0$ constraint
	$$ \|\bsbS\|_{2,0}/p \leq q_g,$$
	it calls for the multivariate version of quantile thresholding $\overrightarrow{\Theta}^\#(\bsbS^T; q_g)$:
	\begin{gather}\label{thres:row}
	\overrightarrow{\Theta}^\#(\bsbS^T; q_g) =
	\begin{cases}
	\bsb0 & \|\tilde{\bsbs}_i\| \leq \lambda_g \\
	\tilde{\bsbs}_i & \|\tilde{\bsbs}_i\| > \lambda_g,
	\end{cases}
	\end{gather}
	where $\lambda_g$ is the $(1-q_g)^{th}$ quantile of the $p$ row norm $\|\tilde{\bsbs}_i\|'s$. The group-wise constraint version can be realized by simply changing $\Theta^\#$ to $\overrightarrow{\Theta}^\#$ in line \ref{Alg1:SUpdate} of Algorithm \ref{Alg:ThetaAlt}, which also satisfies the inequality \eqref{func:suroIneq} given $\rho_k$ properly chosen.

\subsection{Step size }\label{sec:stepsize}

	Define $\tau_k = 1/\rho_k$ as the step size along the gradient in Equation \eqref{func:DefineXi}. To guarantee function value nonincreasing, it suffices to derive a minimal $\rho_k$, or a maximal step size $\tau_k$ such that inequality \eqref{func:suroIneq}  holds. Note that the regularization terms cancel out on both sides of the inequality, hence we focus on the negative log likelihood loss $l(\cdot)$ of $f(\cdot)$ throughout this section. Since the function value of $l(\cdot)$ depends on $\bsba$, $\bsbV$, and $\bsbS$ \textit{only} through $\bsbTh$, we regard $\bsbTh$ as the target variable accordingly.
%	remove the penalty term and rewrite $f(\bsbTh) = -\langle \bsbX, \bsbTh \rangle+ \langle \bsbH, b(\bsbTh) \rangle = l(\bsbTh) $ with some abuse of notation.
%	\begin{gather}
%	h(\bsbTh^{[k]}, \bsbTh^{[k-1]}) = l(\bsbTh^{[k-1]}) + \langle \nabla_{\Theta} l(\bsbTh^{[k-1]}), \bsbDelta \rangle +\frac{\rho_k}{2}\|\bsbDelta\|_F^2,
%	\end{gather}

	Such a $\rho_k$ can be obtained by various line search methods with \eqref{func:suroIneq} served as a stopping criterion. However, in some cases a \textit{universal} step size may be derived based on Taylor expansion. Fortunately $l(\bsbTh)$ is usually smooth under the GLM setting, making Taylor expansion appropriate in order to derive a universal step size.
	
	In the univariate case, for arbitrary $y$ and $x$ and $f$ that is at least differentiable to the second degree, we have the approximation
	$f(y) \approx f(x)+ f'(x)(y-x) + \frac{1}{2}f''(x)(y-x)^2 $.
%	Therefore the multivariate Taylor expansion yields
%	\begin{align*}
%	f(\bsbTh^{[k]}) &= f(\bsbTh^{[k-1]}+\bsbDelta) \\
%	&= f(\bsbTh^{[k-1]}) +\langle \nabla f, \bsbDelta \rangle
%	+\frac{1}{2} Tr\{\bsbDelta^T \bsbH(\bsbTh^{[k-1]}) \bsbDelta\},
%	\end{align*}
%	if the Hessian matrix $\bsbH(\cdot)$ exists, and the Hessian $\bsbH(\cdot) = \bsbH \circ b''(\cdot)$ from the second-order expansion:
	To show the multivariate Taylor expansion for $l(\bsbTh^{[k]})$, we consider a perturbation in the gradient function of the loss:
		\begin{align*}
		\nabla l(\bsbTh^{[k-1]}+\bsbDelta)& = -\bsbX +
		\bsbH \circ b'(\bsbTh^{[k-1]}
		+\bsbDelta)\\
		& = -\bsbX  +\bsbH \circ
		(b'(\bsbTh^{[k-1]})
		+ (\bsbH \circ  b''(\bsbxi^{[k-1]}))\circ
		\bsbDelta \\
		& = \nabla l(\bsbTh^{[k-1]})+
		(\bsbH \circ  b''(\bsbxi^{[k-1]}))\circ
		\bsbDelta ,
		\end{align*}
	with $\bsbDelta = \bsbTh^{[k]}-\bsbTh^{[k-1]}$ for some $ \bsbxi = c \bsbTh^{[k-1]} +(1-c)\bsbTh^{[k]}$ where $c \in (0,1)$.
	Hence
	\begin{align*}
	l(\bsbTh^{[k]}) = & l(\bsbTh^{[k-1]})+
						\langle \nabla_\Theta l(\bsbTh^{[k-1]}), \bsbDelta\rangle+
						\frac{1}{2}\langle\bsbH \circ  b''(\bsbxi^{[k-1]})\circ\bsbDelta, \bsbDelta\rangle.
	\end{align*}

	Denote $\bsbDelta = [\delta_{ij}]$, $\bsbH = [h_{ij}]$, $b''(\bsbxi) = [b_{ij}]$, we have
	
		\begin{align}
		h(\bsbTh^{[k]}, \bsbTh^{[k-1]})- f(\bsbTh^{[k]})
		= 	 & \frac{\rho_k}{2}\|\bsbDelta\|_F^2
		- \frac{1}{2}\langle\bsbH \circ  b''(\bsbxi^{[k-1]})\circ\bsbDelta, \bsbDelta\rangle\\
		\geq & \frac{\rho_k}{2}\|\bsbDelta\|_F^2  -\frac{1}{2} \sum_{i,j} \delta_{ij}^2 h_{ij}b_{ij}\\
		\geq & \frac{\rho_k}{2}\|\bsbDelta\|_F^2 -\frac{1}{2}\sum_{i,j} \delta_{ij}^2 b_{ij}\\
		\geq & \frac{\rho_k}{2}\|\bsbDelta\|_F^2 -\frac{1}{2}\|b''(\bsbxi)\|_{\max} \|\bsbDelta\|_F^2,
		\end{align}	
	where $\|\cdot\|_{\max}$ is defined as the maximal absolute value in the matrix, and $\bsbH$ is a matrix of at most $1's$.	
	
	If $\rho_k \geq  \|b''(\bsbxi)\|_{\max}$, that is, step size $$\tau_k = \frac{1}{\rho_k} \leq 1/\|b''(\bsbxi)\|_{\max},$$  function value will decrease.

	$\|b''(\bsbxi)\|_{\max}$ depends on the specific distribution. Table \ref{tb:functions} provides the forms of $b''(\cdot)$ according to a list of distributions. Notice that it always equals 1 under Gaussian distribution and it is bounded by $\frac{1}{4}$ for Bernoulli, Binomial and Multinomial data. By taking these suprema, nonincreasing function value can be guaranteed via the universal step size choices $\tau = 1$ or $\tau= 4$ respectively for all $k$. 	
	However, in the Poisson, Gamma, and Exponential cases, $\|b''(\bsbxi)\|_{\max}$ has no finite supremum, thus no universal step size may be achieved. Since inequality \eqref{func:suroIneq} only requires decrease on the $(k-1)^{th}$ step, we may select $\tau_k$ based on an ad-hoc reasonably large value of $\|b''(\bsbxi)\|_{\max}$ or from line search methods to satisfy the local descent condition. Since an arbitrary small constant $\tau_k$ easily leads to slow convergence and inaccurate results, line search is vastly preferred for efficient and accurate convergence. The line search scheme will be outlined in details as a part of Algorithm \ref{Alg:Pois_AG2_search}.

%\pagebreak

\subsection{Fast Computation} \label{sec:fast}

	\subsubsection{Accelerated gradient with line search}
	
	The lack of theoretical maximal step size under some distributions is expected to result in slow convergence, which motivates us to find an accelerated algorithm.
	Nesterov's second accelerated first-order method \citep{nesterov1988approach} is a popular tool to improve the convergence speed for unconstrained smooth problem. It can achieve the convergence rate $O(1/k^2)$ where $k$ is the iteration number. This rate is shown to be the optimal convergence rate for smooth and convex first-order problems, and later extended to a large class of non-smooth convex ones including Lasso \citep{beck2009fast}.
	We borrow the framework of accelerated proximal gradient \citep{tseng2008accelerated,parikh2013proximal} and define an operator $\mathcal{P(\cdot)}: \mathbb{R}^n \rightarrow \mathbb{R}^n$ which solves a \textit{non-convex} optimization problem. Theoretical work has been done analyzing the approximation accuracy when such regularization is convex \citep{tseng2010approximation}, to which our problem does not belong. However, experience shows that under non-convex settings the second accelerated method still works well if the step size $\tau_k$ is further reduced and appropriately selected.
	
	For notation simplicity, we take $\bsbTh$ to represent $\bsba$, $\bsbV$ and $\bsbS$ jointly.
	The second method introduces two momentum terms $\bsbY$ and $\bsbnu^{[k]}$ in the $k^{th}$ iterate such that
	\begin{align}
	\bsbY &= (1-\theta_k) \bsbTh^{[k-1]}+ \theta_k \bsbnu^{[k-1]},\\
	\bsbnu^{[k]} & = \mathcal{P}(\bsbTh - \frac{\tau_k}{\theta_k}\nabla l(\bsbY)),\\
	\bsbTh^{[k]} & = (1-\theta_k)\bsbTh^{[k-1]}+\theta_k \bsbnu^{[k]}.
	\end{align}
	$\mathcal{P}$ yields the answers to the non-convex optimization problem
	\begin{gather}
	\begin{split}
	\argmin_{\bsbTh = \bsb1_n\bsba^T + \bsbV \bsbS^T}
	l(\bsbY)+\langle \nabla l(\bsbY), \bsbTh - \bsbY \rangle
	+ \rho_k \theta_k \|\bsbTh - \bsbnu^{[k]}\|_F^2 + P(\bsbS;\lambda)\\			
	\text{subject to } \bsbV^T \bsbV = \bsbI_{r \times r},
	\end{split}
	\end{gather}
	essentially problem \eqref{func:surogate} with $\bsbXi = \bsbTh - \rho_k\theta_k\nabla l(\bsbY)$. $\theta_k$ should be chosen such that $$\frac{1-\theta_k}{\theta_k^2}\leq \frac{1}{\theta_k^2},$$ and $\theta_k = \frac{2}{k+2}$ functions as a simple choice that satisfies this inequality.
	
	When the theoretical step size $\tau$ is not available, it can be relaxed by an initial guess and further decreased in every iterate until the backtracking criterion is met---in our case, inequality \eqref{func:suroIneq}: the idea is to take a conservative step size $\tau^{(0)}$ first in iteration $k$, and repeatedly proceed along the gradient with a subsequently even smaller step size $\tau^{(m)} = \eta \tau^{(m-1)}$ ($0<\eta<1$) until sufficient decrease in function value has been achieved, before entering iteration $k+1$.
%	Under the accelerated proximal settings, theoretical stopping criteria are derived in \cite{tseng2010approximation} with specific reference to inequality (57).
	
	Algorithm \ref{Alg:Pois_AG2_search} lines out \text{\modelname }$^\text{\Lightning}$, geared especially towards the Poisson case. Note that line \ref{prox_start} to line \ref{prox_end} in Algorithm \ref{Alg:Pois_AG2_search} compose the non-convex $\mathcal{P}$ which includes a low-rank and a sparsity regularization.

		\begin{algorithm}[t!]
		\caption{The \text{\modelname }$^\text{\Lightning}$ Algorithm with acceleration and line search}\label{Alg:Pois_AG2_search}
		\textbf{Input}:
		$\bsbX \in \real^{n \times p}$;
		$r$: the desired rank;
		$M_{out}$/$M_{in}$: the maximum outer/inner iteration number; $\varepsilon_{out}/\varepsilon_{in}$: inner and outer error tolerance;
		$\eta (0<\eta <1)$;
		and the initial estimates $\bsba^{[0]} \in \real^{p \times 1}$,
		$\bsbV^{[0]} \in \mathbb{O}^{n \times r}$,
		$\bsbS^{[0]} \in \real^{p \times r}$.
			\begin{algorithmic}[1]
				\State $k \gets 0$;
				\State $\bsbTh^{[0]} \gets \bsb1_n \bsba^{[0]}+\bsbV^{[0]}{\bsbS^{[0]}}^T $;
				\State $\bsbnu^{[0]} \gets \bsbTh^{[0]}$;
				\Repeat
				%			\State $\bsbTh^{[k]}=\bsb1_n {\bsba^{[k]}}^T + \bsbV^{[k]} {\bsbS^{[k]}}^T$, $k = k+1$
				\State $k \gets k+1$;
				\State $\theta_k \gets 1$ when $k = 1,2$, and $\theta_k = \frac{2}{k+2}$ otherwise;
				\State $n_{ls} \gets 0$;
				\State $\bsbnu^{cur}_{ls}\gets\bsbnu^{[k-1]} $, $  \bsbTh^{cur}_{ls}\gets  \bsbTh^{[k-1]}$, $\tau = 1/\|\bsbX\|_{\max}$;								
				\Repeat
				\State $n_{ls} \gets n_{ls} +1$;
				\State $\tau \gets \eta \tau$;
				\State $\bsbY \gets (1-\theta_k) \bsbTh^{cur}_{ls}+ \theta_k \bsbnu^{cur}_{ls}$;
				\State $\bsbnu^{new}_{ls}\gets \bsbnu^{cur}_{ls}-\frac{\tau}{\theta_k}\nabla l_\Theta(\bsbY)$ where $\nabla l_\Theta(\bsbY) = -\bsbX+\bsbH \circ g^{-1}(\bsbY)$; \label{prox_start}
				\State $t \gets 0$;
				\State Initialize $\tilde{\bsba}^{[0]} \gets \bsba^{[k-1]}$,
				$\tilde{\bsbS}^{[0]} \gets \bsbS^{[k-1]}$,
				$\tilde{\bsbV}^{[0]} \gets \bsbV^{[k-1]}$;
				\Repeat
				\State $t \gets t+1$;
				\State $\tilde{\bsba}^{[t]}\gets\frac{1}{n} (\bsbnu^{cur}_{ls} - {\tilde{\bsbV}^{[t-1]}} {\tilde{\bsbS}^{[t-1]}}{}^T)^T \bsb1_n$;
				\State ${\tilde{\bsbS}^{[t]}} \gets \Theta^\#((\bsbnu^{cur}_{ls}{}^T-\tilde{\bsba}^{[t]} \bsb1_n^T){\tilde{\bsbV}^{[t-1]}};q_e)$; \label{Alg2:sparse}
				\State
				\begin{minipage}{\linewidth}
					Compute SVD of $(\bsbnu^{cur}_{ls}-\bsb1_n {\tilde{\bsba}^{[t]}}{}^T) {\tilde{\bsbS}^{[t]}} = \bsbP \bsbD \bsbQ^T$, set ${\tilde{\bsbV}^{[t]}} \gets \bsbP \bsbQ^T$;	
				\end{minipage}
				\Until {$t > M_{in}$ or  changes in $\tilde{\bsba}$, $\tilde{\bsbS}$, $\tilde{\bsbV}$ no bigger than $\varepsilon_{in} $ }
				\State $\bsbnu^{new}_{ls} \gets\bsb1_n \tilde{\bsba}^{[t]}{}^T +{\tilde{\bsbV}^{[t]}} {\tilde{\bsbS}^{[t]}}{}^T$;	\label{prox_end}
				\State $ \bsbTh^{new}_{ls} \gets (1-\theta_k)\bsbTh^{cur}_{ls}+\theta_k \bsbnu^{cur}_{ls} $;			
				\Until	\begin{minipage}{.95\linewidth}
					 {$f(\bsbTh^{new}_{ls}) \leq f(\bsbY)+ \langle \nabla f(\bsbY),\bsbTh^{new}_{ls} - \bsbY\rangle+\frac{\theta_k}{2 \tau}\|\bsbTh^{new}_{ls} - \bsbY\|_F^2 $} or $n_{ls} > 10$\label{Alg2:linesearch}
					\end{minipage}
					\State \begin{varwidth}[t]{\linewidth}
						${\bsba}^{[k]} \gets \tilde{\bsba}^{[t]}$,
						${\bsbS}^{[k]} \gets \tilde{\bsbS}^{[t]}$,
						${\bsbV}^{[k]} \gets \tilde{\bsbV}^{[t]}$;
					\end{varwidth}	
					\State $\bsbnu^{[k]} \gets\bsbnu^{new}_{ls}$, $ \bsbTh^{[k]} \gets \bsbTh^{new}_{ls} $;								
				\Until {$k > M_{out}$ or ($\|\bsbTh^{[k]}-\bsbTh^{[k-1]}\|_{\max} \leq \varepsilon_{out}$ $\&$ $|f^{[k]}-f^{[k-1]}| \leq \varepsilon_{out}$)}\\
				\Return $\bsba^{[k]}$,
				$\bsbV^{[k]}$,
				$\bsbS^{[k]}$.
			\end{algorithmic}
		\end{algorithm}

	\subsubsection{Progressive screening}

	In cases where $p$ is extremely high, it is sometimes just infeasible to iterate till convergence, neither is it suitable to remove a big fraction of dimensions altogether in carrying out the algorithm \textit{once}.
	For instance, $p$ is $1000$ and the desired post-screening dimension is $100$. It is considered greedy to deem 90\% of the original dimensions nuisance from the very first iteration.
	
	To enhance the scalability as well as to reduce such greediness as problem size blows up, it is natural to \textit{adaptively} kill the dimensions depending on the iteration progress. The elimination of one dimension is equivalent to enforcing the row norm of $\bsbS$ to zero. Therefore, by optimizing the problem subject to a row-sparse criterion repeatedly and complying a once-zero-stays-zero strategy, we are able to progressively screen and squeeze the dimensions.
	
	The screening problem is considered under the group-wise sparse setting. Let the desired percentage of nonzero dimensions be $q_g$, then instead of enforcing $q_g $ as the tuning parameter directly in step \ref{Alg2:sparse} of Algorithm \ref{Alg:Pois_AG2_search}, we introduce a sequence $Q(t,k)$ which decreases from $p$ to $q_g p$ and discard the zero dimensions, where $t$ and $k$ stands for the inner and outer loop iteration number respectively. $Q(t,k)/d$ serves as the new sparsity parameter with $d$ denoting the cardinality of the nonzero index set $\mathcal{N}$ in the current iteration. In this way, the problem is tackled in a smaller space as the iteration proceeds. Since reducing the same amount of dimension is often easier at the beginning when the candidate $d$ is much higher than towards the end, a sigmoidal decay best suits the need---$Q(t,k) = 2p/(1+\exp(a T))$ is recommended in particular, where $a \in [0.01, 0.1]$ determines the speed of the decay. $T = k, t, kt$ controls whether the decaying process is dependent on the outer, inner loop progression or both, with the choice of inner loop being the fastest one and outer loop on the conservative side.

	Algorithm \ref{Alg:PS} lines out the progressive screening scheme to replace step \ref{Alg2:sparse} of Algorithm \ref{Alg:Pois_AG2_search}. $\mathcal{N}$ stands for the index set of nonzero dimensions with respect to the original index and it is initialized as  $\mathcal{N}= (1,2,\cdots, p)$.

		\begin{algorithm}[t!]
			\caption{Progressive Screening}\label{Alg:PS}
			\textbf{Input}:
			$\tilde{\bsbS}^{[t]} \in \real^{p \times r}$,
			$	\tilde{\bsba}^{[t]}$, $\bsbX$,
			$Q(t,k)$, $\mathcal{N}$;
		\begin{algorithmic}[1]
				\State $d \gets \card(\mathcal{N})$
				\State ${\tilde{\bsbS}^{[t]}} \gets \overrightarrow{\Theta}^\#(({\bsbnu^{[k]}}^T-\tilde{\bsba}^{[t]} \bsb1_n^T){\tilde{\bsbV}^{[t-1]}};Q(t,k)/d)$;
				\State $\mathcal{J} \gets \{\mathcal{J}: \|{\tilde{\bsbS}^{[t]}}(j,1:r)\|\neq 0\} $;
				\State $\mathcal{N} \gets \mathcal{N}(\mathit{J})$;
				\State $\tilde{\bsbS}^{[t]} \gets \tilde{\bsbS}^{[t]}(\mathcal{J},1:r),
						\bsbX \gets \bsbX(1:n, \mathcal{J}),
						\tilde{\bsba}^{[t]} \gets \tilde{\bsba}^{[t]}(\mathcal{J})	$;		\\
				\Return $\tilde{\bsbS}^{[t]}$,
				$\tilde{\bsba}^{[t]}$,
				$\bsbX, \mathcal{J}$, $\mathcal{N}$.
			\end{algorithmic}
		\end{algorithm}

	The progressive screening design integrated in \text{\modelname } greatly enhances the scalability of our algorithm when problem complexity explodes. Section \ref{ssec:HapMap} includes results with dimension up to $p = 13220$ to demonstrate such scalability when the competing methods take several times longer or even fail to converge within reasonable time.

\section{Numerical Experiment}	\label{sec:experiment}
\subsection{Simulation data} \label{sec:simulation}

	We generate Gaussian, Bernoulli and Poisson simulation data under three different settings. Simulation data set is built similarly to the spiked covariance model, which is popular among sparse PCA related research \citep{johnstone2009consistency, cai2013sparse}. Simulate an $n \times p$ matrix $\bsbX$ such that
	\begin{gather}
	g(E(\bsbX)) = \bsbP \bsbD \bsbQ^T,
	\end{gather}
	where $\bsbP \in \real^{n \times r}$ has independent and identically distributed standard normal entries, $\bsbD$ is a diagonal matrix with $(\lambda_1, \cdots \lambda_r)$ on its diagonal, $\bsbQ$ is a deterministic $p \times r$ orthonormal matrix, and $q^*$ is the true nonzero rate in $\bsbQ$, with $q_e^*$ denoting element-wise nonzero percentage and $q_g^*$ the percentage of nonzero rows.

	The three scenarios are generated with $n = 100$ observations and $p = 200$ dimensions. We further specify the true $r^*$, $q_e^*$ and $q_g^*$ below:
	\begin{inparaenum}[(a)]
		\item $ r^* =1 $ with sparse vector $\bsbq$ such that $q_e^*=\|\bsbq\|_0/p = 1\%$, with the exception of Bernoulli data at $q_e^* = 5\%$;
		\item $r^* = 4$ with element-wise sparse matrix $\bsbQ$ such that $q_e^*=\|\bsbQ\|_0 /(pr)= 8\%$; and
		\item  $r^* = 4$ with row sparse matrix $\bsbQ$ such that $q_g^*= \|\bsbQ\|_{2,0}/p = 20\%$.
	\end{inparaenum}

	We implement logisticPCA \citep{lee2013coordinate} to compare Bernoulli experiments, and sPCA-rSVD \citep{Shen&Huang} is applied for all distribution types.
%	logisticPCA employs an Minimization-Majorization (MM) objective function to calculate a working variable, and approximates the working variable
	
	%% What metrics did we use?
	To gauge the precision of algorithms, we compare the error on  $\bsbTh$, function value, subspace and selection. Different from a supervised problem, the complexity goes up \textit{whenever} $n$ \emph{or} $p$ increases. We thus scale the $\bsbTh$-error to ${1000\|\hat{\bsbTh}-\bsbTh^\star\|_F^2}/{np} $.
%	Deviance is provided as test error, calculated on an individual copy of the training data with independently generated errors.
	Deviance is defined as $2(l(\bsbX;\hat{\bsbTh})-l(\bsbX;\bsbTh_S))$ where $\hat{\bsbTh}$ denotes the estimated parameter and $\bsbTh_S$ that of the saturated model. For simplicity of presentation and comparison, we scale the deviance of \text{\modelname } for each distribution under each setting to 1, and deviance of the other methods as a ratio to the first one.
	In addition, the largest canonical angle between the estimated loading space and the true space is also interesting. Selection accuracy is evaluated by missing rate (MR) which stands for missed nonzero loading elements, and false positive rate (FP) which represents the false alarm rates for actual zero elements.

	100 simulation experiments are run for each distribution under each setting. The 10\% trimmed means of evaluation metrics are given in Table \ref{tb:theSimuTable}, since trimmed means are more robust than means for non-Gaussian metrics and more comprehensive than median for using all the values. Notice that sparsity parameter $q_e$ and $q_g$ differs among settings. $r = 1$ is a much simpler problem compared to the others, thus $q_e = q_e^*$ would suffice. Setting (b) has $q_e = 4q_e^*$ and $q_g= 4q_e^*$ compared to (c) where the sparsity parameter is only twice that of the true nonzero level, because a more conservative sparsity parameter is necessary for keeping the missing rate under control when the loading matrix is simulated element-wise sparse.
	The highly non-convex optimization problem produces local minima, hindering the recovery of the global optimal solution. Therefore we use a multiple-start scheme \citep{rousseeuw1999fast}. It first utilizes $m_1$ random initial points, out of which we choose $m_2$ $(m_2 < m_1)$ points with the lowest function values after $n_1$ ($n_1 = 2$) iterations and proceed until convergence; finally, the initial point producing the lowest function value is selected. Different $m_1$ and $m_2$ are used for the three distributions, since the degree to which non-convexity affects convergence differs across the distributions according to our experiments.
	Notice that for the rank-four scenarios (b) and (c), both the element-wise and group-wise \text{\modelname } are implemented. The element-wise version is employed as a fair reference comparing with the competitive methods due to their sequential property, where any group constraint would act just like an element one. However, the group $\ell_0$ constraint should ideally be utilized for selection and fit under setting (c), and for simplicity and speed under setting (b).

	In the Gaussian experiments, \text{\modelname } almost always demonstrates better results to sPCA-rSVD across all evaluation metrics, establishing a trustworthy ground line for our algorithm. \text{\modelname } also takes a fraction of the computation time of sPCA-rSVD especially when $r^* = 4$, probably due to its \textit{joint} estimating property, as opposed to the rank-by-rank behavior of sPCA-rSVD. In fact, \text{\modelname } always demonstrates higher efficiency across all experiments. Under settings (b) and (c), the group-wise algorithm \text{\modelname$_g$ } further enhances the overall performance compared to the element-wise form, especially in terms of selection accuracy.

	This is also true among the Bernoulli experiments. While our method consistently outperforms the comparison ones across all three settings in every aspect---supporting the superior formulation of the negative log-likelihood serving as the loss function, \text{\modelname$_g$ } is always able to improve the performance even further. Group sparsity constraint is an effective tool unique to our algorithm which helps with fast dimension elimination, makes selection possible and much more accurate. Allowing $q_g$ to take a slightly higher value enables it to function as an upper bound, reducing the missing rate considerably at a small cost of the false positive rate, which in turn improves the space recovery accuracy.
	
	\begin{table}[H]
		\centering	
		\begin{minipage}{\linewidth}
			{\renewcommand{\arraystretch}{0.94}%
				\resizebox{\linewidth}{!}{
					\begin{tabular}{ll ccc cc c}
						\toprule	
						\multirow{2}{*}{Data}	&			
						\multirow{2}{*}{Method}
						& \multicolumn{3}{c}{Error}
						& \multicolumn{2}{c}{Selection}
						& \multirow{2}{*}{Time (s)}\\
						\cmidrule(r){3-5}
						\cmidrule(r){6-7}
						& & $\bsbTh$-Error& Dev & Angle  & MR(\%)	&	FP(\%) \\
						\midrule
						\multicolumn{8}{c}{ \textit{Setting (a): $r^* = 1$, $q_e^* = 0.01$, $q_e = q_e^*$}}\\
						\midrule
						
						\multirow{2}{*}{Gaussian}	
						
						&\text{\modelname }	& 0.39 	& 1.00	& 0.14  &0.00	&0.00	 & 0.23\\
						& sPCA-rSVD			& 0.48 	& 1.17	& 0.14  &0.00	&0.00 	 & 0.25\\
						\vspace{.001 cm}\\		
						
						\multirow{3}{*}{\specialcell{Bernoulli\\ \footnotesize{$q_e^* = 0.05$}}}	
						&\text{\modelname }  & 3.48	&1.00		& 21.17  &15.48	& 0.81	 & 0.58\\
						&logisticPCA  		& 4.00	&1.01 		& 22.19	& 22.00	&1.15 	& 0.48\\
						&sPCA-rSVD 	  		& 4.63	&1.02		& 24.59	& 20.00	&1.05	& 0.51\\
						
						\vspace{.001 cm}\\						
						\multirow{3}{*}{Poisson}			
						&\text{\modelname }$^\text{\Lightning} $
						&  8.83 	&   1.00 &	22.93 	& 49.00	&0.49  & 1.45	\\
						&sPCA-rSVD	      &  49.04	& $-$    &	16.02 	& 15.00	&0.15  & 1.50	\\
						& \text{\modelname }$^{Gau}$
						& 8.74 	& $-$    &	16.02 	& 15.00	&0.15  & 0.36	\\
						\vspace{.001 cm}\\						
						%%%%%%%%%%%%%%%%%%%%%%%%%%%%%%%%%%%%%%%%%%%%%%%%%%%%%%%%%%%%%%%%%%%%%%%%%%%%%%%%%%%%
						\midrule								  	
						\multicolumn{8}{c}{\textit{Setting (b):  $r^* = 4$, $q_e^*  = 0.08$, $q_e = 4q_e^*$, $q_g = 4q_e^*$}}\\
						\midrule
						\multirow{3}{*}{Gaussian}		
						
						&\text{\modelname } 			   & 0.03	& 1.00	& 0.13	& 3.53	&  26.26	& 0.02	\\
						&sPCA-rSVD			   		   & 0.93	& 52.28 & 0.13  & 1.56  &  26.13	& 0.27 \\
						&\text{\modelname }$_g$        & 0.03   & 1.00	& 0.12  & 0.10  &  26.10    & 0.02 \\			  	
						\vspace{.001 cm}\\	
						
						\multirow{4}{*}{Bernoulli}				
						&\text{\modelname }	    & 6.43 &1.00 & 33.22  & 51.33  & 30.55 & 0.79		\\	
						&logisticPCA 		 	& 7.64 &1.03 & 46.38  & 48.70  & 30.32 & 1.41\\
						& sPCA-rSVD				& 9.14 &1.10 & 44.66  & 42.19  & 29.76 & 1.67 \\
						& \text{\modelname }$_g$  & 5.94 &1.01 & 29.53  & 12.29  & 27.16 & 0.79	\\			
						\vspace{.001 cm}\\
						
						\multirow{4}{*}{Poisson}			
						&\text{\modelname }$^\text{\Lightning} $
						&12.31 &1.00  &88.74 &66.17  &31.84	& 1.32\\
						&sPCA-rSVD			&46.35  &$-$  &77.83 &27.64  &27.64   &6.07	\\				
						&\text{\modelname }$^\text{\Lightning}_g $
						&12.25 &0.55  &88.54 &66.26  &31.85 	& 1.29 \\
						&\text{\modelname }$_g^{Gau}$
						&13.09 & $-$  &76.98 &21.31  &27.94   &0.56\\

						\vspace{.001 cm}\\						
						%%%%%%%%%%%%%%%%%%%%%%%%%%%%%%%%%%%%%%%%%%%%%%%%%%%%%%%%%%%%%%%%%%%%%%%%%%%%%%%%%%%%					
						\midrule								  			  	
						\multicolumn{8}{c}{\textit{Setting (c): $r^* = 4$, $q_g^*  = 0.20$, $q_e = 2q_g^*$, $q_g = 2q_g^*$}}\\
						\midrule
						\multirow{3}{*}{Gaussian}			
						&\text{\modelname } 	 & 0.03	 &1.00	& 0.18 	 & 0.19 	& 25.03   &0.19\\ 	
						& sPCA-rSVD             & 0.93  &40.91 & 0.18   & 0.25	    & 25.03   &2.70 \\	
						&\text{\modelname }$_g$ & 0.03  &1.00  & 0.16   & 0.00	    & 25.00   &0.21 \\
						
						\vspace{.001 cm}\\				
						\multirow{3}{*}{Bernoulli}			
						&\text{\modelname } 	 	& 6.05	   & 1.00	& 31.93    & 23.91	& 30.98 & 0.85	\\ 	
						&logisticPCA          	& 7.65	   & 1.03   & 49.33    & 26.64	& 31.66 & 1.52\\
						&sPCA-rSVD 	 		 	& 9.12	   & 1.11 	& 45.18    & 25.00	& 31.25 & 2.51\\
						&\text{\modelname }$_g$  & 5.58     &  1.01	& 28.03	   & 0.15	& 25.04 & 0.79\\
						
						\vspace{.001 cm}\\				
						\multirow{5}{*}{Poisson}			
						&\text{\modelname }$^\text{\Lightning}$	  & 12.31	 &1.00 &88.40	 &66.17  & 31.84 &1.29\\
						&sPCA-rSVD								  & 44.29    & $-$ &84.31 	 &3.20  & 25.80	 &6.49  \\
						&\text{\modelname }$^\text{\Lightning}_g$ & 12.26	 &0.16 &87.97	 &57.06  & 39.27 &1.27\\
						&\text{\modelname }$^{Gau}_g$			  & 14.56	 & $-$   &84.31	 &0.50 & 25.13   &0.60\\
						&{\text{\modelname }$^{Gau}_g$ } 		\tablefootnote{$r = 3r^*$}
						& 11.19	 & $-$   &49.20	 &0.00 &35.71 	&0.69\\								
						\bottomrule
					\end{tabular}
				}
			}
		\end{minipage}	
		\caption{
			\footnotesize
			$n = 100, p = 200$. Trimmed mean (10\%) results of 100 repetitions. Gaussian experiments utilizes 2 out of 10 initial points; Bernoulli 3/20; Poisson examines 5/30. Gaussian methods under the Poisson model produces $\bsbTh$ as estimators of $E(\bsbX)$, and it results in numerically unstable deviance thus omitted in table.
		}\label{tb:theSimuTable}
	\end{table}

	The Poisson simulations are more challenging because of its complexity in step size selection and its numerical tendency to diverge. Algorithm \ref{Alg:Pois_AG2_search} (\text{\modelname }$^\text{\Lightning} $) is applied to all Poisson settings for its comparatively superior performance. Yet the $r = 4$ scenarios still see difficulty in Poisson estimation due to the space ambiguity issues---the set of loading vectors that define a space is not unique. It is suspected that sPCA-rSVD outperforms \text{\modelname } under Poisson because of the simple Gaussian algorithm where numerical issues are less likely to occur. Thus we list our Gaussian algorithm \text{\modelname }$^{Gau}$ for comparison as well. Better results are achieved through the \text{\modelname }$^{Gau}$ and \text{\modelname }$^{Gau}_g$ algorithms. In an attempt to cope with the space ambiguity, \text{\modelname }$^{Gau}_g$ with $r = 3r^*$ is also included under setting (c), because a higher rank parameter introduces more flexibility in defining the subspace. This results in substantially superior space recovery accuracy.
	
	To sum up, our method demonstrates better efficacy and efficiency of the joint space much more accurately under the Gaussian and Bernoulli settings. It is able to directly eliminate nuisance dimensions for either screening or selection purposes due to the group sparsity constraint, regardless of the underlying true sparse pattern of the loading matrix.
	Although the Poisson experiments do not show results as excellent as the other two, we are able to achieve encouraging recovery accuracy with the Gaussian alternative of \text{\modelname }. Relaxation of rank $r$ to a higher number may also help relieve space ambiguity issues. It is important to note that the results and conclusions are restricted to the formulation of the synthetic examples above.

\subsection{SNPs data} \label{ssec:HapMap}
Single nucleotide polymorphism (SNP) data denotes the variation at the level of a single base pair in DNA sequence that occurs at over 1\% within a species. SNPs data attracts a great amount of attention due to the belief of its undiscovered association with various stratification ways, especially diseases. Two tasks are of particular interest: one is to reveal the underlying structure for a certain population, the other is to select particular features meaningful in clustering the observations. Our proposed methodology, while its primary goal is to recover the jointly sparse and low-rank structure, also takes an interest in learning the informative features.

\begin{figure}[H]
	\centering
	\subfigure[The first PC of \text{\text{\modelname }} ]{
		\includegraphics[scale=.29]{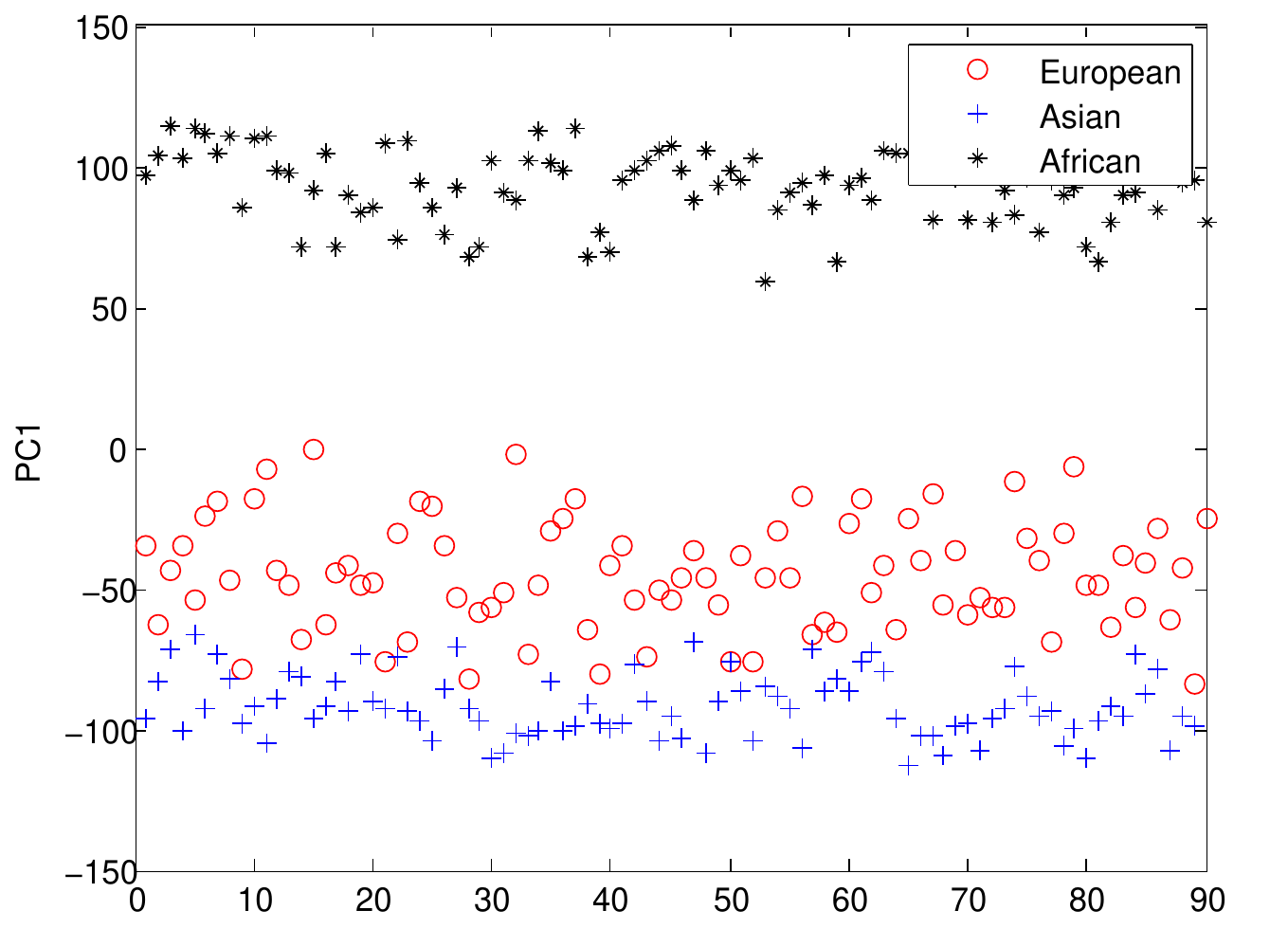}}
	\subfigure[The first two PCs of \text{\text{\modelname }} ]{
		\includegraphics[scale=.29]{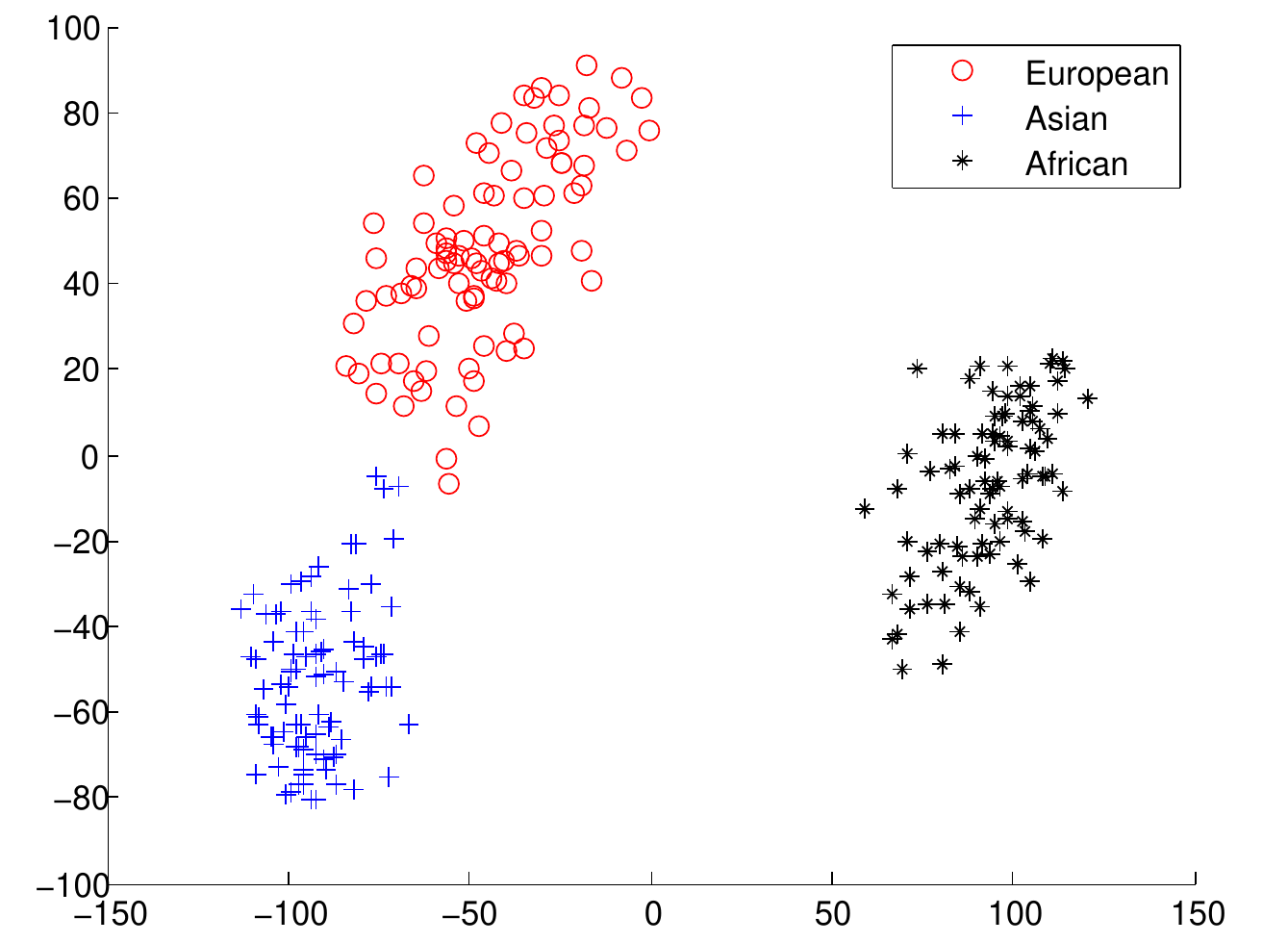}}	
	\subfigure[The first two PCs of \text{\text{\modelname }} ]{
			\includegraphics[scale=.29]{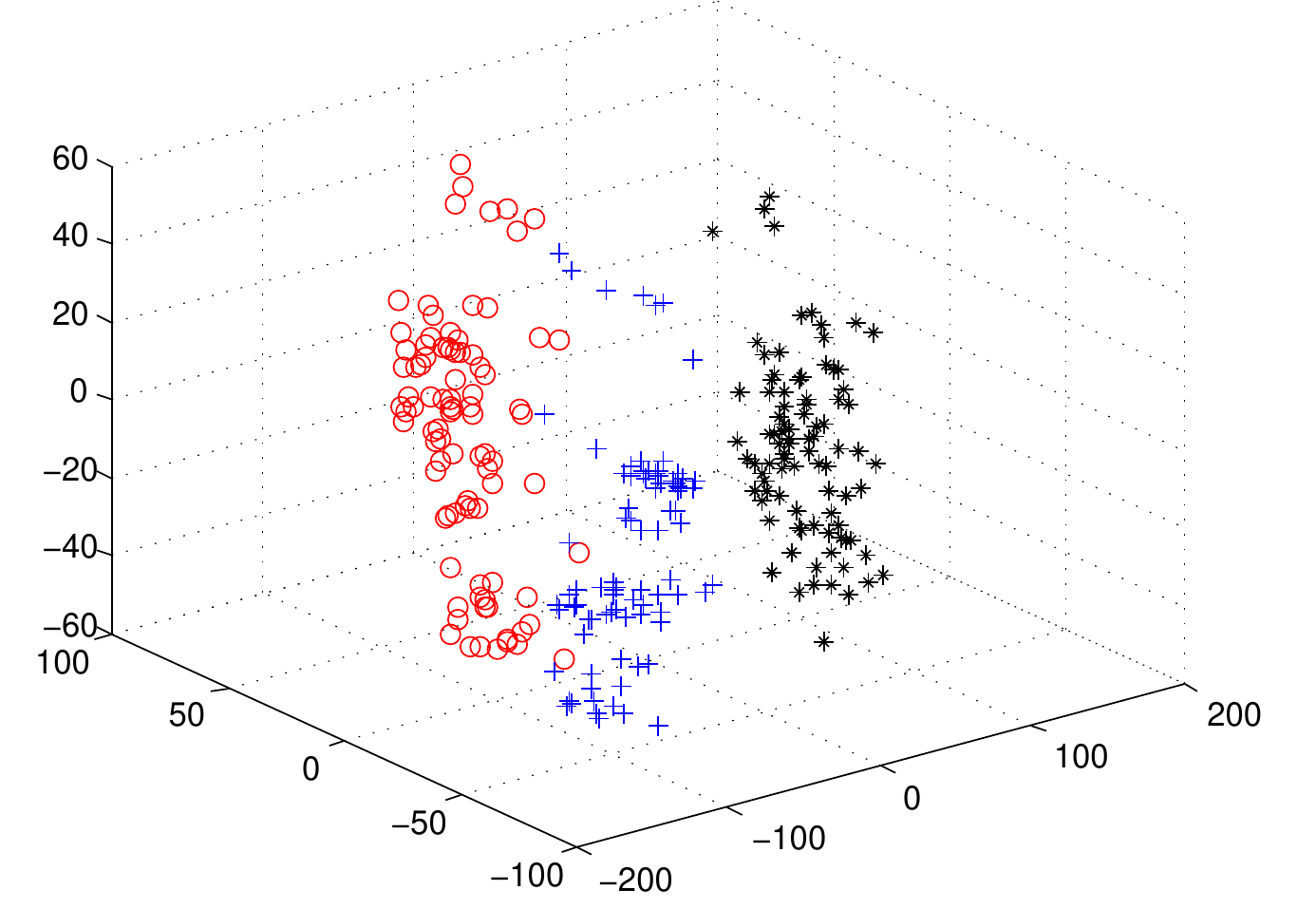}}	
	\caption{\footnotesize \text{\modelname } on HapMap SNPs data. $n = 270$, $p = 1,322$, $r = 3$, $q_g  = 0.10$ with progressive screening, and $q_e = 0.60$ }\label{fg:HapMap Clustering}
\end{figure}

The specific data set we use as an example is the SNPs data made available from the international HapMap project \citep{consortium:2005}. There are 1,322 shared base pair information from 270 observations, consisting of 90 Africans, 90 Caucasians and 90 Asians. The missing rate is 0.53\% and it is handled by the masking scheme. Since the data consist of binary entries denoting the existence of a base pair mutation, traditional dimension reduction with Gaussian assumption is inappropriate. \text{\modelname } is applied with parameters selected as $r=3$, $q_g = 0.10$ and $q_e= 0.60$, while the group-wise sparsity is enforced progressively and the elment-wise one implemented afterwards. Rank and sparsity parameters in our formulation are not as sensitive as some other algorithms such as $\ell_1$. We fix one, alternate the other on a crude scale until the best clustering effect is achieved.
Figure \ref{fg:HapMap Clustering} shows the data projected to the first three principal directions learned by \text{\modelname }. Although the nature of our approach is unsupervised, the European, Asian and African subjects are well separated in the subspace. It is noteworthy that only 12.56\% of the original dimensions are used in all three PCs, out of which only 8 dimensions are shared by the three directions in common.

\begin{figure}[H]
	\centering
	\subfigure[The first PC of \text{\text{\modelname }}]{
		\includegraphics[scale=.29]{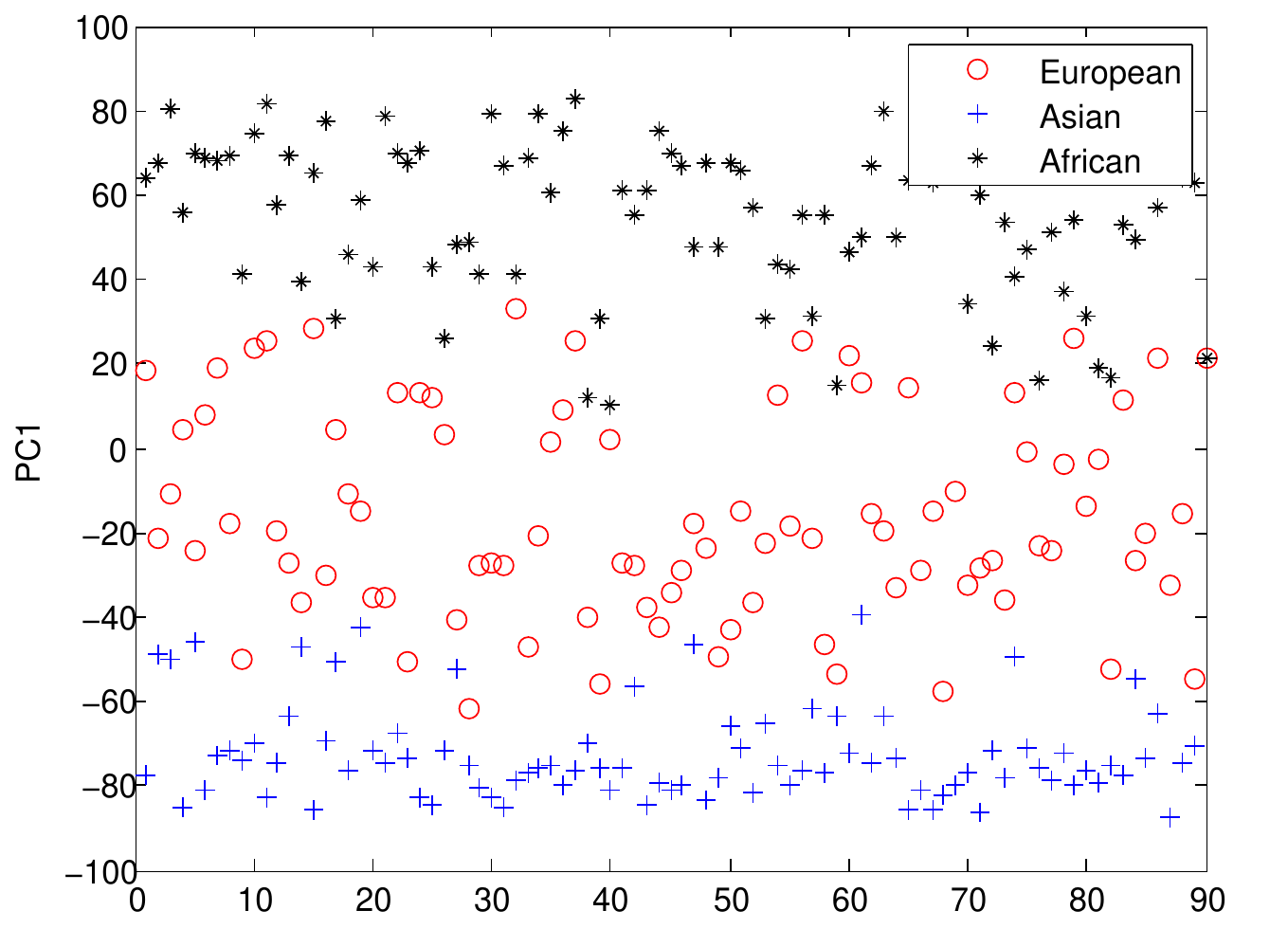}}
	\subfigure[The first two PCs of \text{\text{\modelname }} ]{
		\includegraphics[scale=.29]{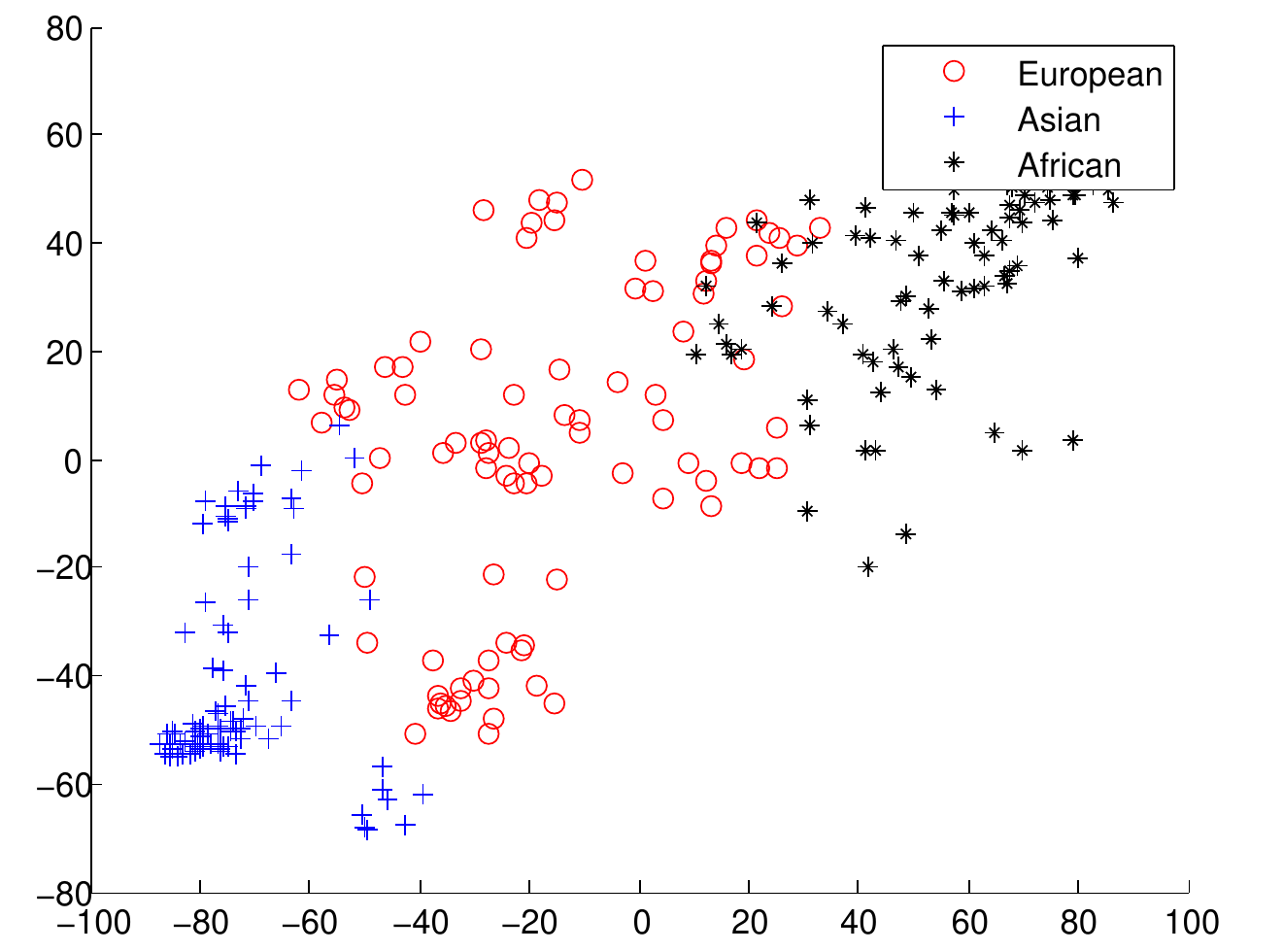}}	
	\subfigure[The first three PCs of \text{\text{\modelname }}]{
		\includegraphics[scale=.29]{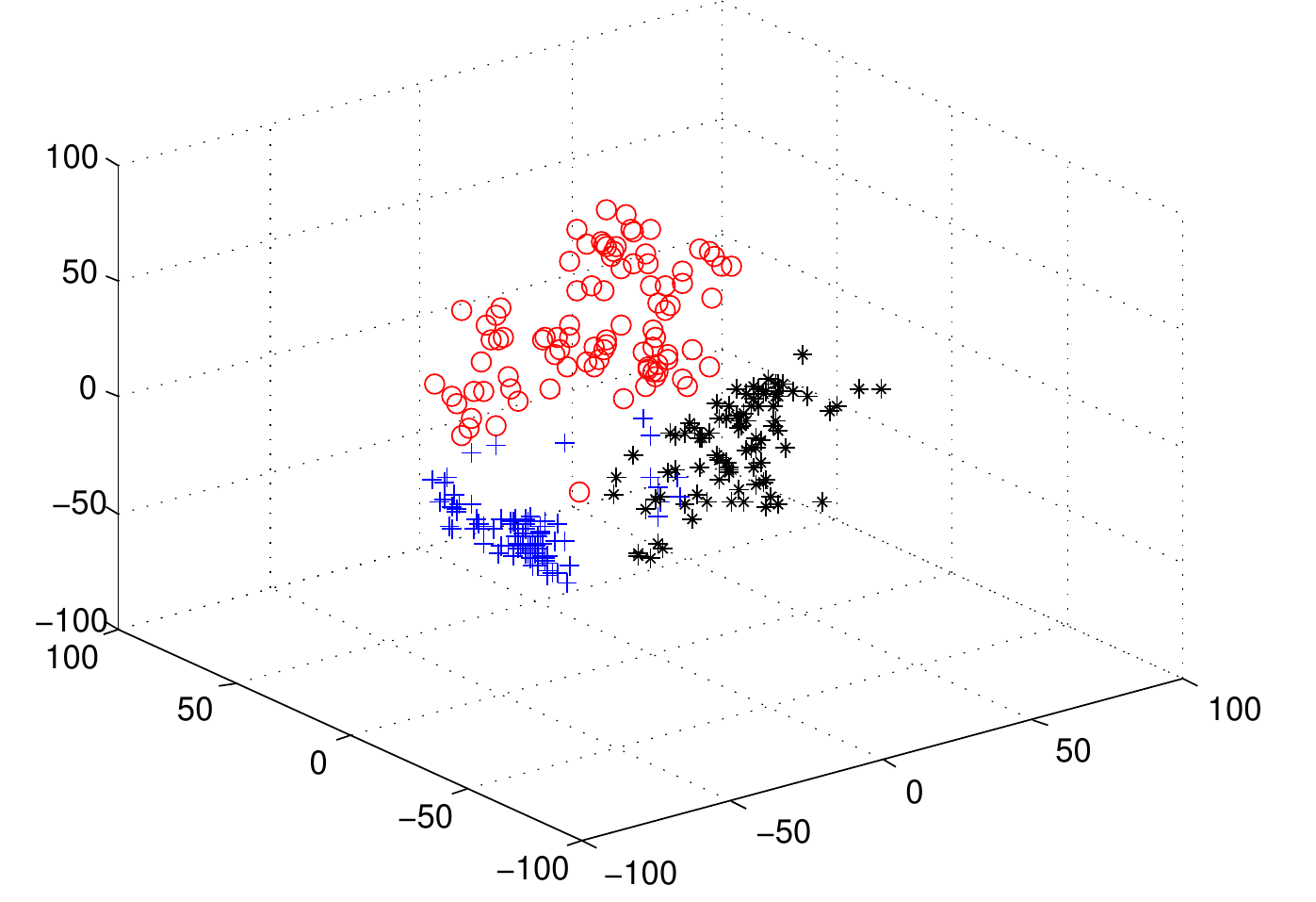}}	
	\caption{\footnotesize \text{\modelname } on inflated HapMap SNPs data. $n = 270$, $p = 13220$, $r = 3$, $q_g  = 0.01$ with progressive screening, and $q_e = 0.60$}\label{fg:HapMap Clustering2}
\end{figure}

To further demonstrate the selection capability and algorithm scalability of \text{\modelname }, we inflate the original 270 by 1,322 data matrix with 9 times as many nuisance dimensions. With parameters chosen at $r = 3$, $q_g = 0.01$ and $q_e = 0.60$, our algorithm is able to produce the clustering in Figure \ref{fg:HapMap Clustering2}. Only 76 are selected from the starting 13220 dimensions and none of the inflated junk dimensions is falsely chosen. It is also worth noting that the entire algorithm with 3 second-stage initial points out of 20 first-stage ones takes 198 seconds, a fraction of the 712 seconds with logisticPCA. sPCA-rSVD appears infeasible under such high-dimensional setting, for each rank takes more than 10 minutes to converge.


%\subsection{A Sequencing data application}
%RNA sequencing data are known to follow the Poisson distribution. To investigate the performance of our method on Poisson data, we use 32000 gene expressions of 14 technical replicates, half of which from a liver sample, the other half from a kidney sample.
%
%%we use RNA sequencing data set of yeast cultures which is available as supplementary materials with \citet{anders2010differential}. The data set consists of 6 observations, each of 7124 genes, 3 replicates per preparation protocols: "random hexamer" (RH) and "oligo(dT)" (dT). These two preparation protocols are treated as two classes.

\subsection{The CNAE-9 text data}

The CNAE-9 text data is a set of 9 categories from the National Classification of Economic Activities. It has been preprocessed so that each document is presented as a row, and each column contains the frequencies of a specific word in the documents.

Three out of nine categories are extracted for use in the experiments, where two thirds of the observations are for training purposes, leaving the remaining one third as testing data sets. The training data consists of 240 observations only, compared to a much larger $p$ of 857.

The data set is highly zero-inflated (99.27\%), which is a native characteristic of text data, as not all words are present in all categories of documents. It is not to be confused with the missing rate 2.8\%, handled by the masking method. The low nonzero entries in data probably further supports the recovery of sparse loadings. A series of progressive group-wise \text{\text{\modelname }}$^\text{\Lightning}$ models are fitted under the Poisson distribution, projecting the data to some lower-dimensional subspaces, after which a K-nearest-neighbor (KNN) classification procedure is carried out on the transformed space to calculate the test misclassification error.

Rank $r$ and sparsity parameter $q_g$ are tuned based on the testing KNN misclassification error. The tuning approach is similar to the HapMap application by a series of experiments, where the parameter values are alternatively changed. Since $q_g$ serves as an upper bound of the true nonzero loading vectors, the result is not very sensitive to $q_g$ as long as a conservative choice is given. Hence an exhaustive grid search is not necessary.

Classification on the reduced dimensions produces considerably better results compared to the original space. Na\"ive KNN on the unprocessed data produces a misclassification rate of 66.67\%---the equivalence of random guess. Regular PCA projects the data to a new coordinate system, greatly enhancing the classification accuracy from an error of 66.67\% to 9.58\% when $r = 8$, and 8.33\% when $r = 20$. However, since no sparse loading vector is pursued, each PC still utilizes all the original words.

\begin{figure}[H]
	\centering
		\includegraphics[scale=.5]{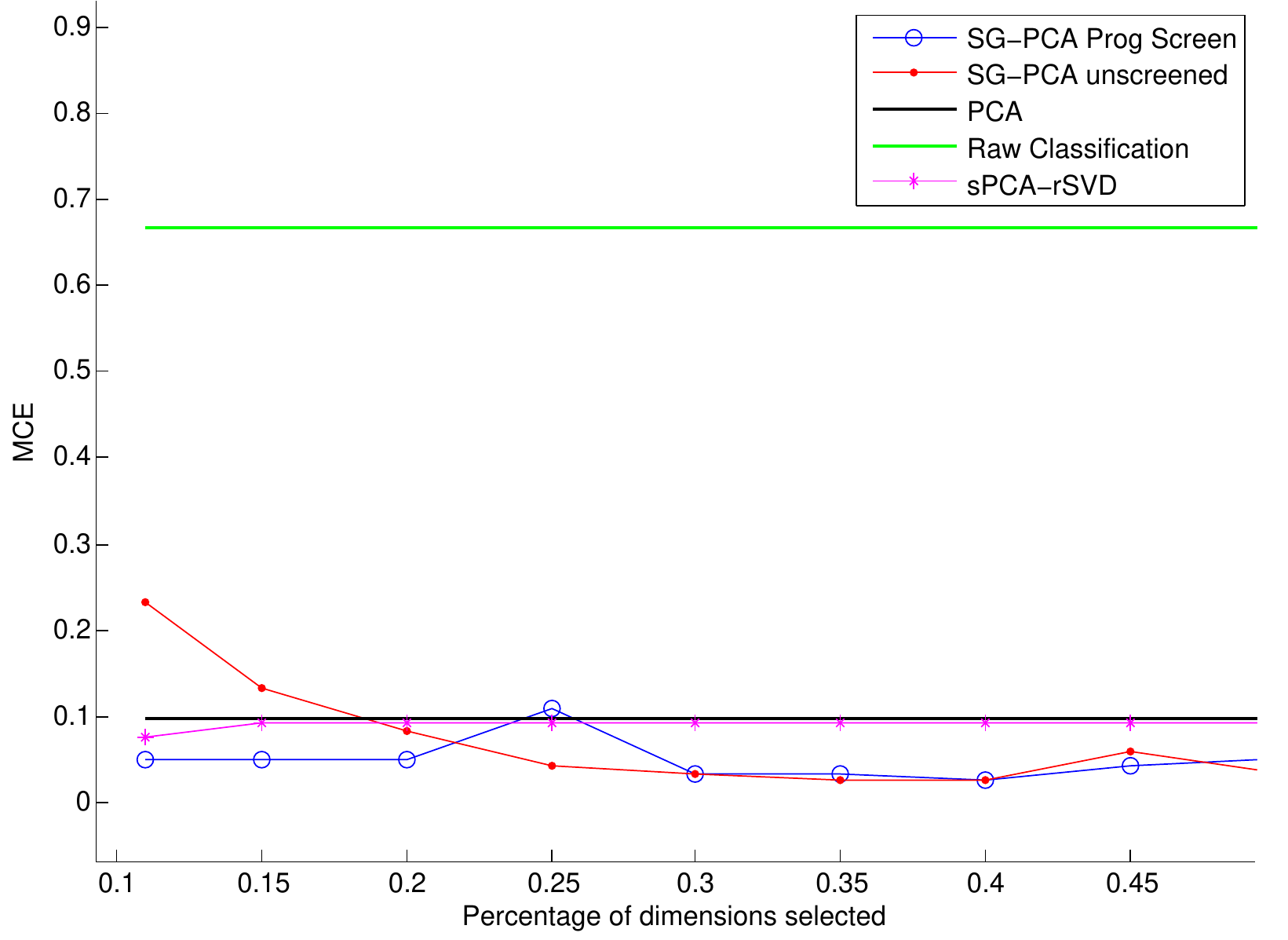}	
	\caption{\footnotesize Misclassification rate comparison of \text{\text{\modelname }}$^\text{\Lightning}$ at r = 8 on CNAE-3 data.}\label{fg:CNAE-9}
\end{figure}

The new loss function coupled with the sparsity constraint delivers a parsimonious representation in the resulting PCs while keeping the strong classification power in the following KNN procedure.
Figure \ref{fg:CNAE-9} shows the misclassification rate at $r = 8$ corresponding to a range of $q_g$ values chosen at $q_g = 0.11, 0.15, 0.20,0.25, 0.30,0.35, 0.40, 0.45,0.50$. In sum, \text{\text{\modelname }}$^\text{\Lightning}$ with progressive screening seem to have the best performance. It surpasses \text{\text{\modelname }}$^\text{\Lightning}$ without progressive screening especially at the lower $q_g$ values because greediness is very much alleviated when the sparsity regularization is stringent. Both achieve a much lower misclassification error than all the other methods, with a selection rate as low as 11\% of the original dimension. sPCA-rSVD performs similarly to regular PCA. \text{\text{\modelname }}$^\text{\Lightning}$ with progressive screening is also the most computationally efficient among all. One multiple-initial-point process consumes on average 12.07 seconds, whereas regular \text{\text{\modelname }}$^\text{\Lightning}$ takes 30.53 seconds and sPCA-rSVD 34.79 seconds to converge.

%\begin{table}
%	\centering
%	\resizebox{\linewidth}{!}{
%	\begin{tabular}{lr lccccr}
%		\toprule
%		\multicolumn{2}{c}{Regular PCA }	&\multicolumn{6}{c}{\text{\modelname }$^\text{\Lightning}$ ($r = 8$)} \\
%		\cmidrule(r){1-2} \cmidrule(r){3-8}
%		$r = 8$	& $r = 20$ &
%		 $q_g = 0.03$      &     $q_g = 0.05$      &     $q_g = 0.08$  &     $q_g = 0.11 $   &  $q_g = 0.15$ &     $q_g = 0.20$  \\
%		\midrule
%		9.58\%	& 8.33\% 	&  18.75\%&    10.42 \%   &  6.94\%&    5.56\%&    7.64\%    &  7.64\%\\
%		
%		\bottomrule
%	\end{tabular}	
%	}
%	\caption{\footnotesize CNAE text data mis-classification rate comparison}\label{tb:CNAE9}
%	
%\end{table}

\section{Summary}\label{sec:summary}

	\text{\modelname } algorithm designs a scalable distribution-specific methodology for the purpose of unsupervised low-rank and sparse data representation, such that PCA is both rectified in theory and accurate to exponential family distributions beyond Gaussianity.
%	It can easily be combined with supervised techniques for regression and classification problems.
	Missing values are taken into consideration with almost no cost to computation efficiency. Nesterov's second accelerated gradient method as well as line search are incorporated for faster optimization	especially under circumstances where theoretical maximal step size is not calculable. A progressive screening strategy is employed to alleviate the greedy nature of sharp dimension reduction, while enhancing the scalability.
	Although the {canonical} link function is favored throughout the paper for its convenience, non-canonical links can also be handled with the same surrogate function technique, as long as the inverse link functions are differentiable.
	
%	A frequent issue with Poisson data in particular is the underestimation of its variance. 	Many remedies have been proposed such as quasi-likelihood approaches, Negative Binomial distribution, or the employment of an extra random effect term.	
	
	However, the convenience of masking the missing values is based upon the assumption that the observations are independent given the missing indices. In fact, the whole model depends on the \textit{conditional} independence assumption among the observations---apart from the association made possible by the low-rank projection of the multivariate data matrix. Such an assumption coincides with that of the regular PCA, but it is obviously an over-simplification of reality. An immediate extension is to corporate association structure into formulating the model, capturing correlation among the non-Gaussian covariates. %This is an under-going research topic.

%\section*{Acknowledgement}
%%The authors would like to thank the editor and referee for their careful comments and useful suggestions that improve the quality of the paper.
%This work was supported in part by NSF grant DMS-1352259.

%	The novel adoption of manifold method in sparse PCA problems have proved to associate with increased optimality compared to sequential and batch methods at an affordable sarifice of computation cost while the other two remain as alternative options.

%\bibliographystyle{abbrv}
\bibliographystyle{plainnat}
\bibliography{SGPCA_Lynn}

\end{document}